\documentclass[10pt,a4paper]{article}
\usepackage{hyperref}
\usepackage[english]{babel}
\usepackage[utf8]{inputenc}
\usepackage[T1]{fontenc}
\usepackage{graphicx}
\usepackage{geometry}
\usepackage{tabularx}
\usepackage{psfrag}
\usepackage{fancyhdr}
\usepackage{xcolor}
\usepackage{sistyle}
\usepackage{amsfonts}
\usepackage{amsmath}
\usepackage{amssymb}
\usepackage{booktabs}
\usepackage{multirow}
\usepackage{amsthm}
\usepackage{subcaption}
\usepackage{pifont}
\usepackage[ruled]{algorithm2e}
\usepackage{enumerate}
\usepackage{enumitem}
\usepackage{bbold}
\usepackage{tikz}
\usepackage{hyperref}
\usepackage{mathrsfs}
\usepackage{xr}
\graphicspath{{figures/}}

\definecolor{carmin}{rgb}{0.7294,0.0392,0.0392}

\theoremstyle{theorem}
\newtheorem{proposition}{Proposition}
\newtheorem{corollary}{Corollary}

\newtheorem{lemma}{Lemma}
\newtheorem{theorem}{Theorem}
\theoremstyle{definition}
\newtheorem{definition}{Definition}

\theoremstyle{remark}
\newtheorem{remark}{Remark}

\def\bxi{\boldsymbol{\xi}}
\def\e{\mathrm{e}}
\def\d{\mathrm{d}}

\begin{document}

\title{A covariant, discrete time-frequency representation \\tailored for zero-based signal detection
\thanks{The authors acknowledge support from ERC grant Blackjack (ERC-2019-STG-851866) and ANR AI chair Baccarat (ANR-20-CHIA-0002), and thank Julien Flamant and Adrien Hardy for insightful discussions.}}

\author{Barbara Pascal, Rémi Bardenet.
\thanks{B. Pascal and R. Bardenet are with Univ. Lille, CNRS, Centrale Lille, UMR 9189 CRIStAL, F-59000 Lille, France (e-mail: barbara.pascal@univ-lille.fr, remi.bardenet@gmail.com).}}

\maketitle

\begin{abstract}
Recent work in time-frequency analysis proposed to switch the focus from the maxima of the spectrogram toward its zeros, which, for signals corrupted by Gaussian noise, form a random point pattern with a very stable structure leveraged by modern spatial statistics tools to perform component disentanglement and signal detection.
The major bottlenecks of this approach are the discretization of the Short-Time Fourier Transform and the boundedness of the time-frequency observation window deteriorating the estimation of summary statistics of the zeros, on which signal processing procedures rely.
To circumvent these limitations, we introduce the \emph{Kravchuk transform}, a generalized time-frequency representation suited to \textit{discrete} signals, providing a covariant and numerically tractable counterpart to a recently proposed discrete transform, with a \textit{compact} phase space, particularly amenable to spatial statistics.
Interesting properties of the Kravchuk transform are demonstrated, among which covariance under the action of $\mathrm{SO}(3)$ and invertibility.
We further show that the point process of the zeros of the Kravchuk transform of white Gaussian noise coincides with those of the spherical Gaussian Analytic Function, implying its invariance under isometries of the sphere.
Elaborating on this theorem, we develop a procedure for signal detection based on the spatial statistics of the zeros of the Kravchuk spectrogram, whose statistical power  is assessed by intensive numerical simulations, and compares favorably to state-of-the-art zeros-based detection procedures.
Furthermore it appears to be particularly robust to both low signal-to-noise ratio and small number of samples.
\end{abstract}

\tableofcontents

\section{Introduction}

 \textbf{Context.} Time-frequency analysis is the most adapted tool to describe and process nonstationary signals, due to its ability to simultaneously capture events that are localized in time and a dynamically evolving frequency content.
Among the many known representations~\cite{flandrin1998time}, the spectrogram, defined as the squared modulus of the short-time Fourier transform, is one of the most natural.
It provides a natural energy distribution in the time-frequency plane, the maxima of which correspond to the presence of information of interest.
Thus, the precise localization of the maxima of the spectrogram has been thoroughly studied, leading to the development of sophisticated techniques such as \textit{ridge extraction}, \textit{reassignment} and \textit{synchrosqueezing}~\cite{flandrin1998time,flandrin2018explorations} to name but a few, which can be leveraged to perform demodulation of real signals~\cite{meignen2017demodulation}.

From another point of view, it has recently been remarked that the zeros of random spectrograms, seen as a random point pattern in the time-frequency plane, possess a peculiarly regular structure~\cite{gardner2006sparse,flandrin2015time}.
This opened a dual perspective on time-frequency analysis, shifting the interest from spectrogram maxima toward the zeros of the spectrogram, which rather reflect the absence of signal.
One intuition in favor of considering the zeros rather than maxima is that,  for a broad range of noise levels in the data, including high noise
, the zeros show a rigid spatial organization, while the structure of the maxima intrinsically lacks robustness to noise and deformations, thus requiring heavy procedures~\cite{meignen2017demodulation}.

 \textbf{Related work.} Observing that the zeros of the spectrogram tend to repel each other and spread uniformly all over the time-frequency plane at the only exclusion of the region where the underlying deterministic signal lies, \cite{flandrin2015time} proposed a filtering procedure relying on the identification of abnormal distances between close-by zeros, which \cite{bardenet2020zeros} modified into a more direct identification of holes in the pattern of zeros.
A similar methodology has been adapted to the Paul-Daubechies Continuous Wavelet Transform~\cite{abreu2018filtering, bardenet2021time}. 
The motivation of \cite{abreu2018filtering} comes from filtering audio signals using the zeros of their scalograms, while the authors of \cite{bardenet2021time} establish the common theoretical ground on which zero-based time-frequency processing is based, demonstrating the close connection between representations of the complex white Gaussian noise and particular \textit{Gaussian Analytic Functions} (GAFs).
A GAF is a type of random function that is analytic on a domain of the complex plane.
These random functions have recently caught the attention of the probability community~\cite{hough2009zeros}.
GAFs lie behind many signal processing theoretical results, though in an implicit way, such as in the pioneering work~\cite{flandrin2015time}.
Their explicit identification~\cite{bardenet2020zeros,bardenet2021time} motivates a systematic investigation of analytic-valued signal representations.

In particular, identifying the distribution of the zeros of the spectrogram of white noise and the zeros of the so-called \emph{planar} Gaussian Analytic Function, \cite{bardenet2020zeros} developed statistical tests for signal detection that rely on the properties of the zeros of that particular Gaussian Analytic Function.
Considering data of the form
\begin{align}
\label{eq:def_data}
\boldsymbol{y} =  \mathsf{snr} \times \boldsymbol{x} + \boldsymbol{\xi},
\end{align}
where $\boldsymbol{x}$ is a deterministic signal of interest corrupted by complex white Gaussian noise $\boldsymbol{\xi}$, with $\mathsf{snr} \geq 0$ the signal-to-noise ratio, signal detection consists in determining, given an observation $\boldsymbol{y}$, whether there is a such a non-zero signal of interest, $\boldsymbol{x}\neq 0$ and 
$\mathsf{snr} >0$, or whether $\boldsymbol{y}$ consists in pure noise.
Such a task has been a long-standing problem in statistics~\cite[Chapter 10]{wasserman2004all}, with numerous applications in signal processing, ranging from radar~\cite{gashinova2010signal} to finance~\cite{chang2011dynamic} and astrophysics~\cite{chassande1999time,abbott2016observation}. 
\cite{flandrin2015time,bardenet2020zeros,abreu2018filtering} all use spatial statistics tools to design detection tests in the setting \eqref{eq:def_data}.
We also note that, recently, (non-zero) level sets of the spectrogram have also been investigated for the detection of elementary Hermite functions, with theoretical guarantees on the performance of the test~\cite{ghosh2021estimation}.

Another key link with GAFs is the introduction in \cite{bardenet2021time} 
of transforms on $\mathbb{C}^{N+1}$ based on discrete orthogonal polynomials, which map white Gaussian noise to the so-called \emph{spherical} GAF. 
Our work is a direct continuation of that line.

 \textbf{Goals, contributions and outline.}
There are two bottlenecks to developing procedures based on the zeros of the standard spectrogram. 
First, the continuous Fourier transforms involved need to be approximated by discretization. 
Implicitly, this requires tuning the width of the analysis window, which amounts to set the time-frequency resolution; see the discussion in~\cite[Section 5.1.2]{bardenet2020zeros}.
A good tuning requires prior knowledge about the characteristic time and frequency scales of the underlying signal, which can be inaccessible in practice.
Moreover, the effect of approximating the continuous Fourier transforms in the Fourier spectrogram on the existence and extraction of zeros are largely unknown.
Second, in practice, only a bounded window in the time-frequency plane is observed. 
The accurate estimation of functional statistics of the pattern of zeros thus requires sophisticated edge corrections~\cite{moller2003statistical}.

Initially looking for a time-frequency interpretation of either one of the discrete transforms introduced in~\cite[Section 4.5]{bardenet2021time}, we draw here inspiration from the physical literature on coherent states to construct a novel discrete time-frequency transform.
Unlike the transforms in \cite[Section 4.5]{bardenet2021time}, our transform has no hyperparameter like a window width. 
Moreover, unlike the Short-Time Fourier transform, the phase space associated with this new transform is compact, and the transform of white noise almost surely has $N$ zeros.
This drastically simplifies the estimation of spatial statistics.

Like the transform of \cite[Section 4.5]{bardenet2021time}, when applied to a standard white Gaussian vector, the zeros of our Kravchuk spectrogram have the distribution of the zeros of the so-called \emph{spherical} Gaussian analytic function, a well-known random polynomial.
This kind of similarity with \cite[Section 4.5]{bardenet2021time} is more than chance: we shall actually see that, up to a stereographic change of variables, our transform is the product of a non-vanishing, non-analytic prefactor with the transform introduced in \cite[Section 4.5]{bardenet2021time}.
From a signal processing point of view, the prefactor is key, though.
Indeed, we show that, unlike related discrete transforms motivated by orthogonal polynomial arguments in~\cite[Section 4.5]{bardenet2021time},
our Kravchuk transform possesses the traditional properties of a time-frequency representation, such as covariance and a \textit{resolution of the identity}, providing stable reconstruction.
In addition, connecting the discrete transform of \cite[Section 4.5]{bardenet2021time} to the proposed Kravchuk transform provides a weak 
time-frequency interpretation of the former.
This echoes our initial motivation, while leaving open the precise correspondence between the spherical and the time-frequency phase spaces.

An alternative formulation of our transform enables us to provide a numerically stable scheme for the computation of the corresponding spectrogram, and a robust  algorithm for extracting its zeros.
Then, using spatial statistics on the sphere, we propose a detection procedure based on the zeros of our new spectrogram, along the lines of \cite{bardenet2020zeros}.
We give exhaustive empirical evidence that the resulting detection test is more robust to high noise levels and low sample size than the tests based on the zeros of the Short-Time Fourier transform with Gaussian window of \cite{bardenet2020zeros}.

Section~\ref{sec:stateofart} reviews the key steps followed by~\cite{gardner2006sparse,flandrin2015time,bardenet2020zeros,bardenet2021time}, from standard time-frequency analysis to the description of the zeros of the Fourier spectrogram of complex white Gaussian noise as zeros of a Gaussian Analytic Function.
Our new covariant discrete transform is designed in Section~\ref{ssec:def_transform}, and its main properties are listed.
A direct characterization of the zeros of our spectrogram is derived in Section~\ref{ssec:zeros_spherical}.
Practical implementation is discussed in detail in Section~\ref{sec:implementation}.
Finally, the detection procedure based on the zeros of the novel spectrogram is developed in Section~\ref{sec:detect}, and assessed by numerical experiments exploring a wide range of situations in Section~\ref{sec:exp}.

\begin{figure*}
\centering
\begin{subfigure}{0.19\linewidth}
\includegraphics[width = \linewidth]{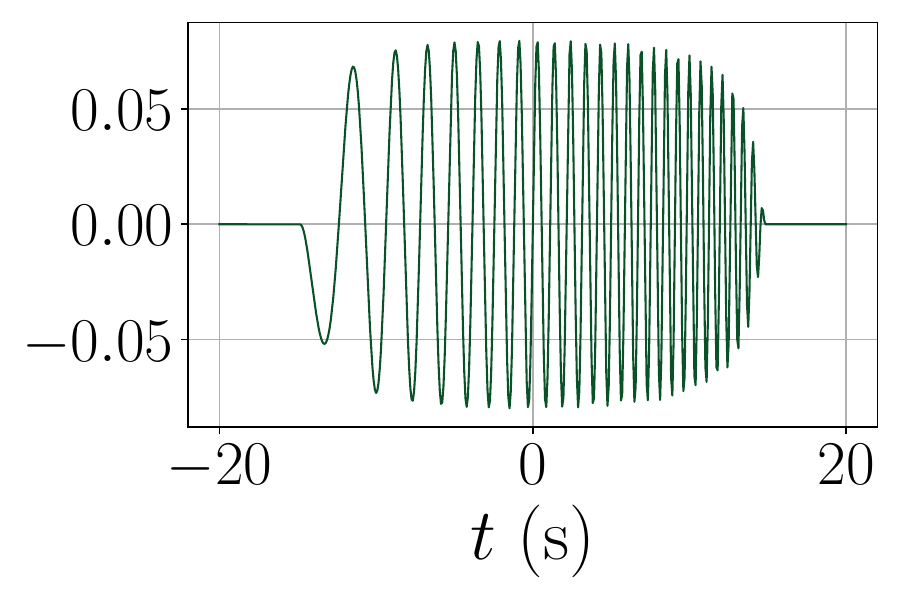}
\subcaption{$\mathsf{snr} = \infty$}
\end{subfigure}
\begin{subfigure}{0.19\linewidth}
\includegraphics[width = \linewidth]{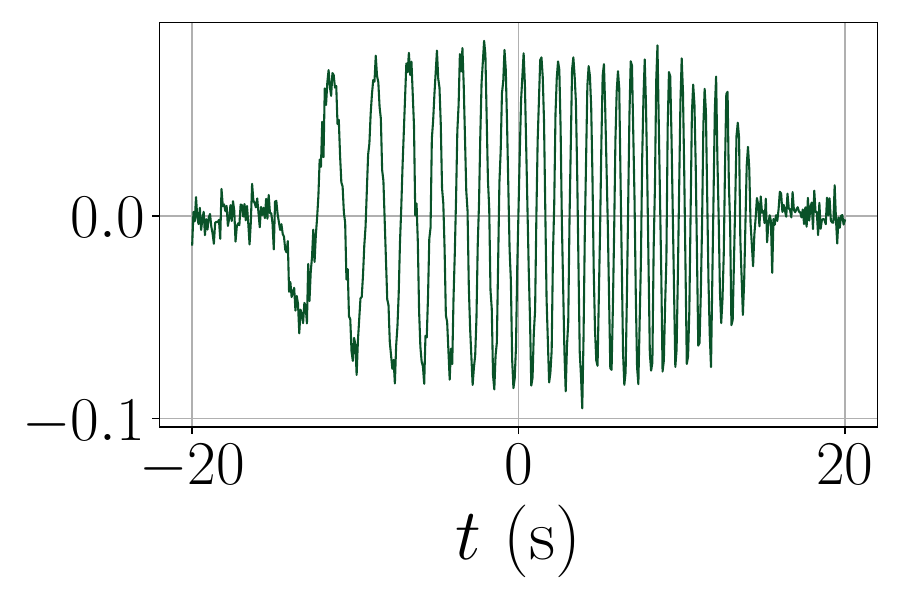}
\subcaption{$\mathsf{snr} = 5$}
\end{subfigure}
\begin{subfigure}{0.19\linewidth}
\includegraphics[width = \linewidth]{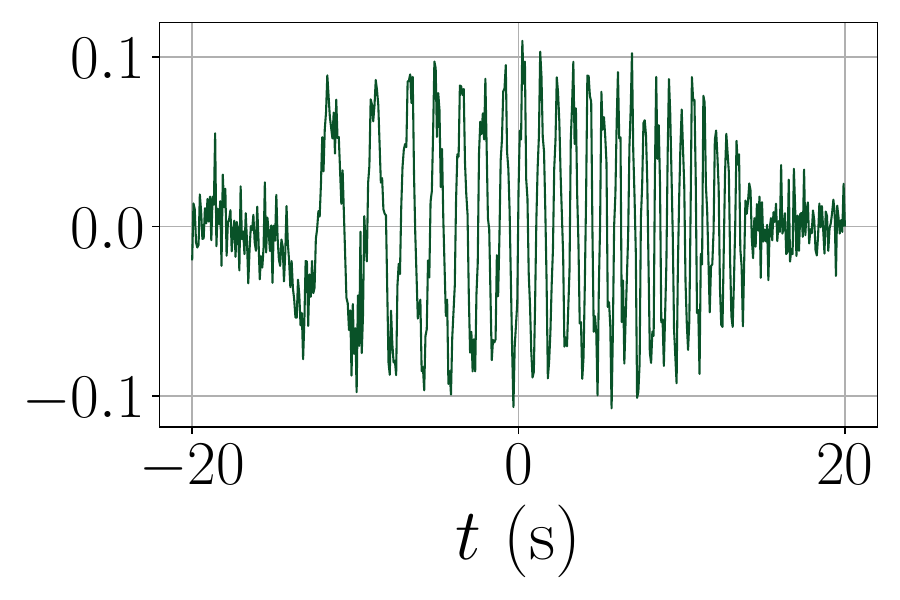}
\subcaption{$\mathsf{snr} = 2$}
\end{subfigure}
\begin{subfigure}{0.19\linewidth}
\includegraphics[width = \linewidth]{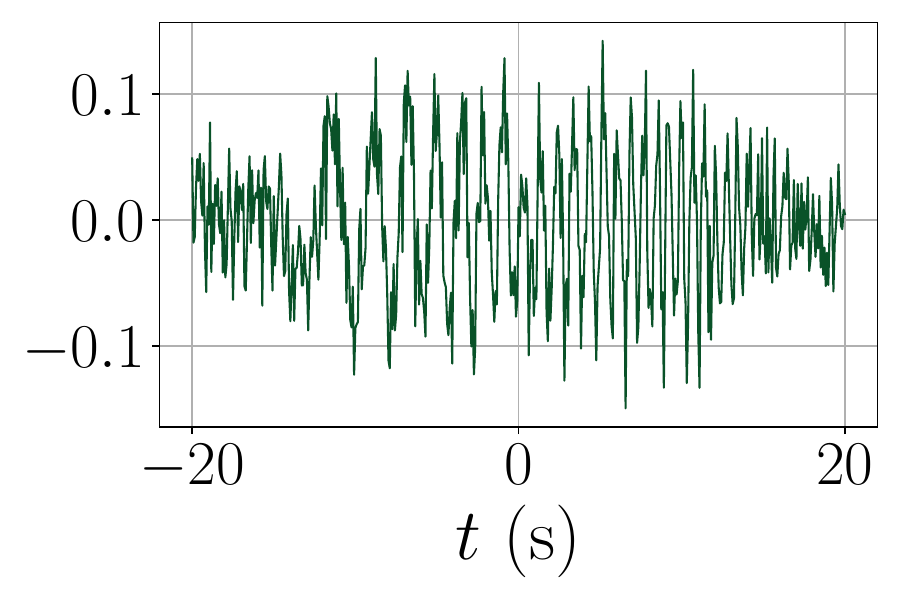}
\subcaption{$\mathsf{snr} = 1$}
\end{subfigure}
\begin{subfigure}{0.19\linewidth}
\includegraphics[width = \linewidth]{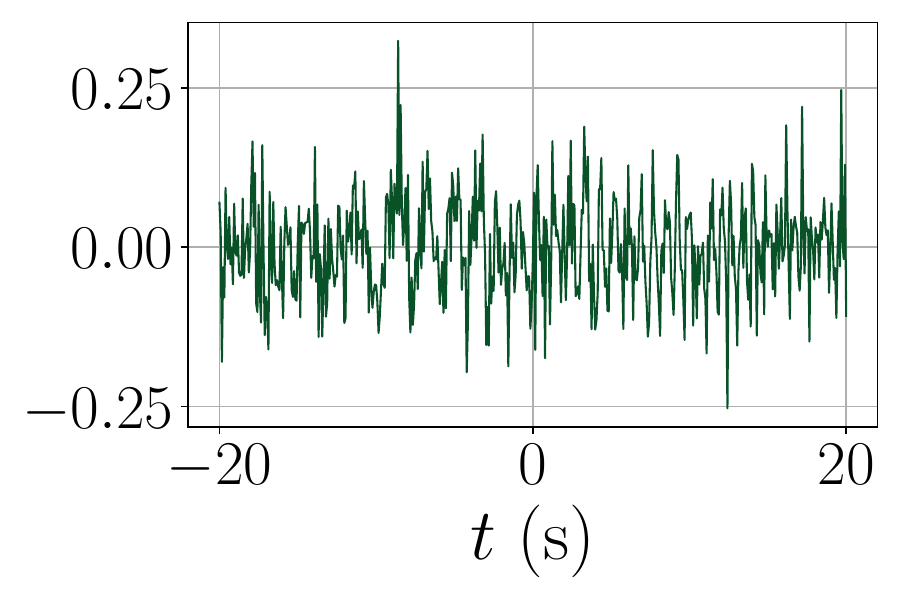}
\subcaption{$\mathsf{snr} = 0.5$}
\end{subfigure}
\caption{\label{fig:nchirp}\textbf{Chirp signals immersed in white noise~\eqref{eq:def_data}.} Deterministic chirp~\eqref{eq:def_chirp} of duration $2\nu = 30$~s,  with characteristic frequencies $f_1 = 0.5 $~Hz and $f_2 = 1.25$~Hz, observed for $40$~s and embedded in complex white Gaussian noise,  with $N+1 = 513$ sample points. The signal-to-noise ratio $\mathsf{snr}$ decreases from left to right.}
\end{figure*}

 \textbf{A typical waveform.} 
Many real-world signals, \textit{e.g.},  gravitational waves~\cite{innocent1997wavelets} or ultrasound recording of bats~\cite{auger2013time}, are well described by \textit{chirps}, consisting in waveforms of limited duration modulated in amplitude and frequency. 
A widely used parametric model is
\begin{align}
\label{eq:def_chirp}
x(t) = A_{\nu}(t) \times \sin \left( 2\pi \left(f_1 + (f_2-f_1)\frac{(t+\nu)}{2\nu} \right)t \right),
\end{align}
where the  time-varying instantaneous frequency increases linearly from $f_1$ at time $-\nu$ to $f_2$ at time $\nu$, and 
$A_{\nu}(t)$ is an infinitely differentiable function with compact support $[-\nu, \nu]$. 
Figure~\ref{fig:nchirp} presents examples of noisy observations following~\eqref{eq:def_data}, where the deterministic signal is of the form~\eqref{eq:def_chirp}, for different noise levels.
For the sake of illustration, we shall systematically illustrate both standard tools and our contributions on signals following Model~\eqref{eq:def_chirp}.
Note however that the procedures we introduce are nonparametric, and thus by no means restricted to chirps.

 \textbf{Notations.} 
The group of rotations of $\mathbb{R}^3$ is defined as $\mathrm{SO}(3) = \lbrace \boldsymbol{R} \in \mathbb{R}^{3\times 3}, \, \, \boldsymbol{R}^\top \boldsymbol{R} = \boldsymbol{I}, \, \mathrm{det}(\boldsymbol{R}) = 1\rbrace$, where $^\top$ denotes the matrix transpose, $\boldsymbol{I}$ is the identity matrix and $\mathrm{det}$ the determinant of a matrix.
Complex-valued functions of the real variable $t$ are denoted $y(t)$.
Defining $\lVert y \rVert_2 = \int_{\mathbb{R}} \lvert y(t) \rvert^2 \mathrm{d}t$, $L^2(\mathbb{R}) = \lbrace y : \mathbb{R} \rightarrow \mathbb{C}, \,  \lVert y \rVert_2 < \infty \rbrace$ is the space of \textit{finite energy} signals.
For $N \in \mathbb{N}$,  $\boldsymbol{\mu} \in \mathbb{C}^{N+1}$ and $\boldsymbol{C} \in \mathbb{R}^{(N+1) \times (N+1)}$, $\mathcal{N}_{\mathbb{C}}(\boldsymbol{\mu},\boldsymbol{C})$ denotes the Gaussian vector of mean $\boldsymbol{\mu}$ and covariance matrix $\boldsymbol{C}$.
Discrete signals, obtained, \textit{e.g.}, by sampling a function $y$ of $\mathbb{R}$ at $N+1$ points, are stored as column vectors $\boldsymbol{y} = \left( \boldsymbol{y}[\ell] \right)_{\ell = 0}^N$, with, \textit{e.g.}, $\boldsymbol{y}[\ell] = y(t_{\ell})$ the $\ell^{\text{th}}$ sample. Finally, $\overline{\boldsymbol{y}}$ denotes the entrywise complex conjugates of $\boldsymbol{y}$.

\section{Zeros of the standard Fourier spectrogram}
\label{sec:stateofart}

\subsection{From time-frequency analysis to the Bargmann transform}

Given a short-time window $h\in L^2(\mathbb{R})$, either having a compact support or decreasing fast outside of a bounded interval, the \textit{Short-Time Fourier Transform} of a signal $y\in L^2(\mathbb{R})$ consists in the decomposition of the signal over the family of time-translated and frequency-modulated replica of $h$~\cite{grochenig2001foundations},
\begin{align}
\label{eq:def_STFT}
V_h y(t, \omega) = \int_{-\infty}^{\infty} \overline{y(u)} h(u - t) \mathrm{e}^{ -\mathrm{i}  \omega u } \, \mathrm{d}u.
\end{align}
The \textit{Fourier spectrogram} is then defined as the squared modulus of the Short-Time Fourier Transform and, provided that $\Vert h\Vert_2 =1 $, it satisfies
\begin{align}
\label{eq:cons_energy}
\int \int_{\mathbb{R}} \left \lvert V_h y(t, \omega) \right\rvert^2 \, \mathrm{d}t \dfrac{\mathrm{d}\omega}{2\pi} = \lVert y \rVert_2^2.
\end{align}
The Fourier spectrogram is thus often interpreted as a time-frequency energy distribution~\cite{flandrin1998time,grochenig2001foundations,flandrin2018explorations}.
Furthermore, the energy conservation of Equation~\eqref{eq:cons_energy} comes with reconstruction formul\ae~\cite[Section~3.2]{grochenig2001foundations}, which are crucial to perform, \textit{e.g.}, component separation~\cite{flandrin2015time}. 

When it comes to the study of the zeros of the Fourier spectrogram, the choice of a circular Gaussian analysis window, $g(t) = \pi^{-1/4}\mathrm{e}^{-t^2/2}$, is common~\cite{flandrin2015time,bardenet2020zeros}, since it is essentially the only window providing an analytic transform\footnote{On spectrogram zeros and non-Gaussian windows, see \cite[Theorem 1.9]{haimi2020zeros}.}~\cite{ascensi2009model}.
Indeed, introducing $z = (\omega + \mathrm{i}t)/\sqrt{2}$, the Gaussian Short-Time Fourier Transform coincides, up to a nonvanishing function, with the Bargmann transform~\cite[Chapter~3]
{grochenig2001foundations}
\begin{align}
\label{eq:def_Barg}
\forall z \in \mathbb{C}, \quad By(z) = \frac{ \mathrm{e}^{ - z^2/2}}{\pi^{1/4}} \int_{\mathbb{R}} \overline{y(t)} \mathrm{e}^{ \sqrt{2}tz - t^2/2 } \, \mathrm{d}t,
\end{align}
\textit{via} the relation
\begin{align}
\label{eq:STFT-Bargmann}
V_g y(t,\omega) = \mathrm{e}^{ - \lvert z\rvert^2/2}  \mathrm{e}^{ -\mathrm{i}\omega t /2} B y(z).
\end{align}
First introduced in quantum physics~\cite{bargmann1961hilbert} as an interlacing operator between the Schrödinger and the Fock representations, the Bargmann transform caught afterward the attention of the signal processing community~\cite{daubechies1988frames} due its ability to provide analytic representations of signals.
In particular, the analyticity of the Bargmann transform of $y\in L^2(\mathbb{R})$ ensures that its zeros are isolated points of the complex plane.
Intuitively, identifying the time-frequency plane to the complex plane through $z = (\omega + \mathrm{i}t)/\sqrt{2}$, the zeros of the Fourier spectrogram of a noisy signal can then be seen as a random configuration of points in the complex plane, and can thus be analyzed with the tools of spatial statistics~\cite{flandrin2015time,bardenet2020zeros}.

\begin{figure*}
\centering
\begin{subfigure}{0.19\linewidth}
\includegraphics[width = \linewidth]{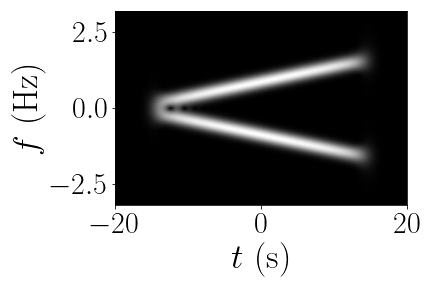}
\subcaption{$\mathsf{snr} = \infty$}
\end{subfigure}
\begin{subfigure}{0.19\linewidth}
\includegraphics[width = \linewidth]{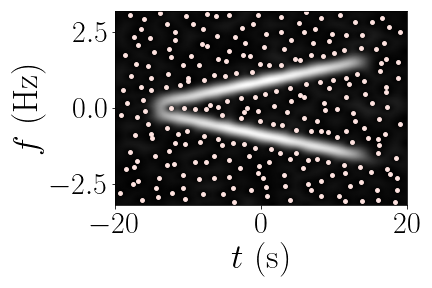}
\subcaption{$\mathsf{snr} = 5$}
\end{subfigure}
\begin{subfigure}{0.19\linewidth}
\includegraphics[width = \linewidth]{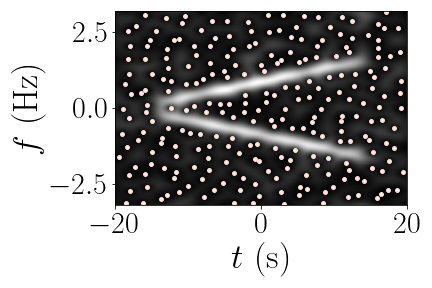}
\subcaption{$\mathsf{snr} = 2$}
\end{subfigure}
\begin{subfigure}{0.19\linewidth}
\includegraphics[width = \linewidth]{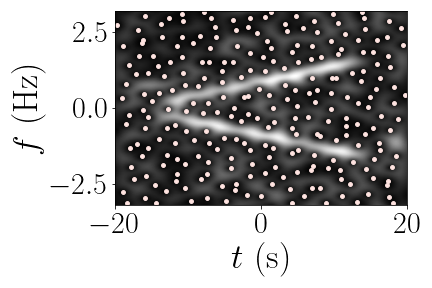}
\subcaption{$\mathsf{snr} = 1$}
\end{subfigure}
\begin{subfigure}{0.19\linewidth}
\includegraphics[width = \linewidth]{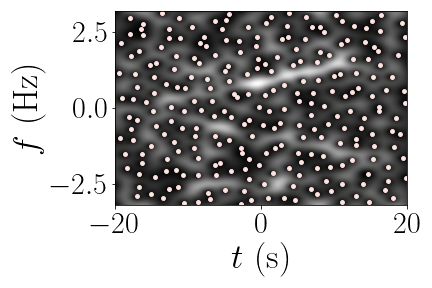}
\subcaption{\label{sfig:nstft_05}$\mathsf{snr} = 0.5$}
\end{subfigure}
\caption{\label{fig:nstft}\textbf{Fourier spectrogram of noisy chirps.} Squared modulus of the Gaussian Short-Time Fourier Transform of the signals of Figure~\ref{fig:nchirp}, with zeros indicated by pale rose dots. The signal-to-noise ratio $\mathsf{snr}$ is decreasing from left to right.}
\end{figure*}

\subsection{Zeros of the  spectrogram of complex white Gaussian noise}
\label{ssec:zeros_tf}

Fourier spectrograms of the noisy chirps of Figure~\ref{fig:nchirp} are displayed in Figure~\ref{fig:nstft}, with their zeros indicated by pale rose dots.
As can be observed in Figure~\ref{sfig:nstft_05}, when observations are dominated by noise, the zeros are evenly spread,  while larger and larger holes in the zeros pattern appears at the location of the signal in the time-frequency plane as the signal-to-noise increases.
Detection procedures developed by~\cite{flandrin2015time,bardenet2020zeros} rely on the measurement of the discrepancy between the observed configuration of zeros and the reference situation of pure noise.
Because white Gaussian noise does not correspond to a signal in $L^2(\mathbb{R})$, there was a need to rigorously characterize the distribution of the zeros of the Fourier spectrogram of white noise.

\begin{figure}
\centering
\begin{subfigure}{0.49\linewidth}
\centering
\includegraphics[height = 2.5cm]{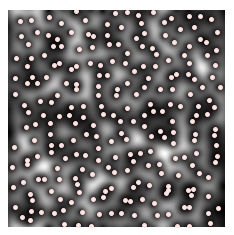}
\subcaption{\label{sfig:GAFC}planar $\mathsf{GAF}_{\mathbb{C}}$}
\end{subfigure}
\begin{subfigure}{0.49\linewidth}
\centering
\includegraphics[height = 2.5cm]{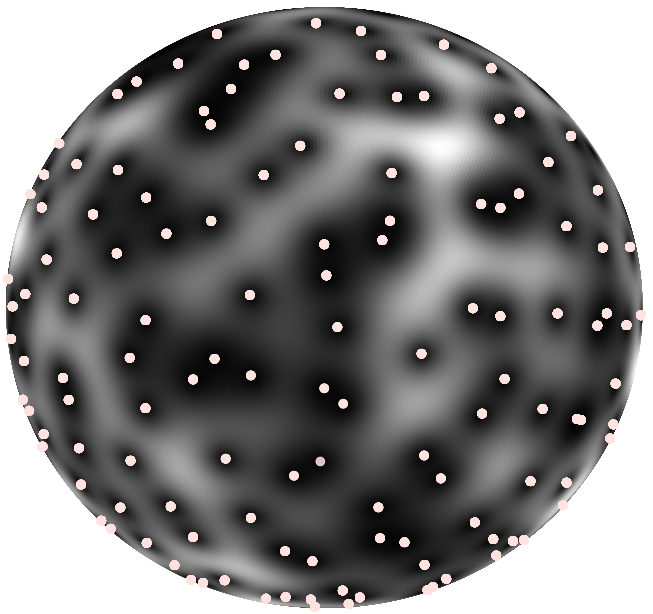}
\subcaption{\label{sfig:GAFS}spherical $\mathsf{GAF}_{\mathbb{S}}$}
\end{subfigure}
\caption{\textbf{Gaussian Analytic Functions.} In grey level: squared modulus of the planar and spherical Gaussian Analytic Functions, respectively introduced at Equations~\eqref{eq:defGAFC}~and~\eqref{eq:defGAFS}, in their natural geometry, pale rose dots indicates their zeros.}
\end{figure}

The first step toward characterization of the zeros \cite{bardenet2020zeros,bardenet2021time} is to expand complex white Gaussian noise onto the Hilbertian basis of $L^2(\mathbb{R})$ formed by Hermite functions $\lbrace h_k, \, k = 0,1, \hdots\rbrace$. 
The latter functions have a very simple closed-form Bargmann transform~\cite[Section 3.4]{grochenig2001foundations}, namely $B h_k(z) = z^k/\sqrt{k!}$.
Then, using linearity and carefully studing the convergence of the series, one can compute the Bargmann transform of white noise $\xi = \sum_{k\in \mathbb{N}} \langle \xi, h_k \rangle h_k$
\begin{align}
\label{eq:Barg_noise}
B \xi(z) = \sum_{k = 0}^{\infty} \langle \xi, h_k\rangle \frac{z^k}{\sqrt{k!}}.
\end{align} 
The probabilist's eye then recognizes the so-called \emph{planar Gaussian Analytic Function}
\begin{align}
\label{eq:defGAFC}
\mathsf{GAF}_{\mathbb{C}}(z) = \sum_{n=0}^{\infty} \xi[n] \frac{z^n}{n!}, \quad \xi[n] \sim\mathcal{N}_{\mathbb{C}}(0,1) \, \, \text{i.i.d.},
\end{align}
whose modulus is displayed in grey level in Figure~\ref{sfig:GAFC}, its zeros being indicated by pale rose dots.
In particular, the zeros of the spectrogram of white Gaussian noise coincide in law with the zeros of the planar Gaussian Analytic Function.
The latter distribution has been fully characterized; see ~\cite[Section~3.4]{hough2009zeros}.
Notably, the distribution of zeros is invariant under isometries of the plane, as can be observed from Figure~\ref{sfig:GAFC}. 
This invariance is of primary importance in the construction and estimation of the summary statistics used in detection tests~\cite{bardenet2020zeros}.

\subsection{Algebraic interpretation and the covariance principle}
\label{ssec:algebra_CS}

The invariance under isometries of the plane of the zeros is deeply linked to a core property of the time-frequency representation~\eqref{eq:def_STFT}: its covariance with respect to time and frequency shifts~\cite{grochenig2001foundations,flandrin2018explorations}.
Covariance properties of representations is a major topic in the theory of signal processing~\cite{cohen1995time}, and has been widely documented, notably in the cases of the Short-Time Fourier Transform~\cite[Chapter~9]{grochenig2001foundations} and of the Continuous Wavelet Transform~\cite{ali2000coherent}, establishing a fruitful bridge between quantum physics~\cite{gazeau2009coherent,perelomov2012generalized} and signal processing~\cite{torresani1999two}. 
This original perspective, consisting in the identification of an underlying symmetry group, not only provides precious insights on the properties of signal representations~\cite{cohen1995time,boashash2015time}, but also yields general alternative formulations~\cite{torresani1999two}, and can be exploited in applications, as illustrated by gravitational wave detection~\cite{innocent1997wavelets}.
Importantly for us, it has been shown that, given a symmetry group, one can construct a covariant representation~\cite{perelomov2012generalized}, known as the \textit{coherent state} decomposition, with completeness properties.
Before taking advantage of this algrebraic framework to design a novel transform at Section~\ref{sec:novel}, we briefly describe the construction of the Short-Time Fourier transform through the \textit{Weyl-Heisenberg group}.

From a Hilbert space point of view, the Short-Time Fourier Transform of a signal  can be interpreted as the scalar product
\begin{align}
V_h y(t,\omega) = \langle y,  \boldsymbol{W}_{(t, \omega)} h \rangle, \quad  \boldsymbol{W}_{(t, \omega)} h (u) = e^{-\mathrm{i} \omega u} h(u-t)
\end{align}
between the signal and a family of functions $\lbrace  \boldsymbol{W}_{(t, \omega)} h, \, (t,\omega) \in \mathbb{R}^2 \rbrace$, called coherent states, and obtained by applying time translations and frequency modulations to the analysis window $h$.
The operators $\boldsymbol{W}_{(t,\omega)}$ act unitarily and transitively on $L^2(\mathbb{R})$, satisfy the non-commutative composition rule
\begin{align}
\label{eq:Weyl-Heisenberg}
 \boldsymbol{W}_{(t', \omega')}  \boldsymbol{W}_{(t, \omega)} = \mathrm{e}^{\mathrm{i}\omega t'}  \boldsymbol{W}_{(t+t', \omega+\omega')}.
\end{align}
Then, the operator family $\lbrace \mathrm{e}^{\mathrm{i}\gamma} \boldsymbol{W}_{(t,\omega)}, \, (\gamma, t,\omega) \in [0, 2\pi] \times\mathbb{R}^2\rbrace$ 
constitutes the Weyl-Heisenberg group, whose group law derives from~\eqref{eq:Weyl-Heisenberg}.
By construction, up to a pure phase factor, the family of coherent states $\lbrace  \boldsymbol{W}_{(t, \omega)} h, \, (t,\omega) \in \mathbb{R}^2 \rbrace$ is invariant under the action of the Weyl-Heisenberg group.
The reconstruction formula for the Short-Time Fourier Transform is equivalent to the \textit{overcompleteness} of the coherent state family~\cite[Chapter 9]{grochenig2001foundations}, \textit{i.e.}, a signal can be exactly reconstructed from the knowledge of its inner products with all the coherent states.
Finally, the covariance of the Short-Time Fourier Transform under the time-frequency shifts, writes, for any signal $y\in L^2(\mathbb{R})$,
\begin{align}
V_h [\boldsymbol{W}_{(t,\omega)} y](t', \omega') = \mathrm{e}^{-\mathrm{i}(\omega'-\omega)t} V_h y(t'-t,\omega'-\omega),
\end{align}
involving an extra phase term, which disappears when taking the squared modulus to obtain the spectrogram.
In particular, the covariance of the Fourier spectrogram 
under time-frequency shifts
ensures that the performance of an algorithm relying on spectrograms does not depend on the \textit{a priori} unknown location of the signal in the time-frequency plane.

\section{A new covariant discrete transform}
\label{sec:novel}

The purpose of this section is to construct a novel covariant representation, specifically designed for \textit{discrete} signals, in order to circumvent both the theoretical difficulty of defining continuous white noise~\cite[Section 3.1]{bardenet2020zeros} and \cite[Section 3.2]{bardenet2021time}, and the subtle practical question of discretizing the Short-Time Fourier transform~\cite[Section 5.1]{bardenet2020zeros}.
To that end, we consider the algebraic framework of Section~\ref{ssec:algebra_CS}, and choose as underlying symmetry group the group of rotations $\mathrm{SO}(3)$. 
This group acts irreducibly on the finite-dimensional space $\mathbb{C}^{N+1}$ of digital signals, $N \in \mathbb{N}$.
Then, inspired by the physics literature on coherent states~\cite{gazeau2009coherent,perelomov2012generalized}, we introduce what we call the \textit{Kravchuk transform}, derive its main properties.
Finally, we study the distribution of the zeros of the associated \textit{Kravchuk spectrogram}.

\subsection{Definition of the Krachuk transform}
\label{ssec:def_transform}

\subsubsection{The Kravchuk basis} 
\label{sssec:krav_basis}
The first step is to identify the orthonormal basis in which the Kravchuk transform has a comprehensible explicit expression.
Following~\cite{arecchi1972atomic,gazeau2009coherent}, this basis is built from the \textit{symmetric Kravchuk polynomials}, consisting in a collection of $N+1$ polynomials, which are orthogonal with respect to the symmetric binomial measure of parameter, $1/2$ and the associated $N+1$ Kravchuk functions.
Denoting by $Q_n(t;N)$ the evaluation at $t$ of the Kravchuk polynomial of order $n$ associated to the symmetric binomial measure with $N$ trials, then the orthogonality relation writes
\begin{align}
\label{eq:ortho_poly}
\sum_{\ell = 0}^N \binom{N}{\ell} Q_n(\ell;N) Q_{n'}(\ell;N) = 2^N \binom{N}{n}^{-1} \delta_{n,n'},
\end{align}
where $\delta_{n,n'}$ denotes Kronecker's delta.
Defining the Kravchuk \textit{functions} as
\begin{align}
\label{eq:def_Krav_fun}
q_n(\ell;N) = \frac{1}{\sqrt{2^N}}  \sqrt{\binom{N}{n}} Q_n(\ell;N) \sqrt{\binom{N}{\ell}},
\end{align}
and the associated column vectors $\boldsymbol{q}_n = \left( q_n(\ell;N) \right)_{\ell = 0}^N$, which will be abusively called Kravchuk functions as well in the following, \eqref{eq:ortho_poly} induces that the family $\lbrace \boldsymbol{q}_n, \, n = 0,1, \hdots, N\rbrace$ is an orthonormal basis of $\mathbb{C}^{N+1}$, namely the \textit{Kravchuk basis}.

\subsubsection{Decomposition into $\mathrm{SO}(3)$ coherent states} 

Adapting the decomposition onto the family of $\mathrm{SO}(3)$ coherent states from quantum physics~\cite[Chapter~6]{ali2000coherent} to the framework of signal processing and discrete signals, leads to the following definition of a novel covariant representation.

\begin{definition}
\label{def:krav_trans}
For a discrete signal $\boldsymbol{y} \in \mathbb{C}^{N+1}$,  the generalized covariant time-frequency transform (or simply \emph{Kravchuk transform}) of $\boldsymbol{y}$ is
\begin{align}
\label{eq:expr_K}
T\boldsymbol{y}(\vartheta, \varphi) =   \sum_{n = 0}^N \sqrt{\binom{N}{n}} \left( \cos \frac{\vartheta}{2} \right)^{n} \left( \sin \frac{\vartheta}{2} \right)^{N - n} \mathrm{e}^{ \mathrm{i} n \varphi } (\textbf{Q}\boldsymbol{y})[n],
\end{align}
where $(\vartheta,\phi) \in [0,\pi]\times [0, 2\pi]$ are the spherical coordinates parameterizing the phase space $S^2$, and 
\begin{align}
\label{eq:decomp_Krav}
(\textbf{Q}\boldsymbol{y})[n] = \langle \boldsymbol{y}, \boldsymbol{q}_n \rangle= \sum_{\ell = 0}^N \overline{\boldsymbol{y}[\ell]} q_n(\ell;N) \quad
\end{align}
are the coefficients of the vector $\boldsymbol{y}$ in the orthonormal basis of Kravchuk functions $\lbrace \boldsymbol{q}_n, n = 0, 1, \ldots, N \rbrace$, seen as vectors $\left( q_n(\ell;N)\right)_{\ell = 0}^N$ with $N+1$ points.
\end{definition}

We remark that the Kravchuk transform \eqref{eq:expr_K} naturally embeds in the algebraic framework presented in Section~\ref{ssec:algebra_CS} in the case of the Short-Time Fourier transform. 
Indeed, consider the vectors
\begin{align}
\label{eq:SO3_CS}
\boldsymbol{\Psi}_{\vartheta, \varphi} = \sum_{n = 0}^N \sqrt{\binom{N}{n}} \left( \cos \frac{\vartheta}{2} \right)^{n} \left( \sin \frac{\vartheta}{2} \right)^{N - n} \mathrm{e}^{ \mathrm{i} n \varphi } \boldsymbol{q}_n,
\end{align}
for $\vartheta \in [0,\pi]$ and  $\varphi \in [0, 2\pi]$.
By construction,  $T\boldsymbol{y}(\vartheta, \varphi) = \langle \boldsymbol{y}, \boldsymbol{\Psi}_{\vartheta, \varphi} \rangle.$
As we shall see shortly, Proposition~\ref{prop:kravchuk} then ensures that the family of vectors introduced in~\eqref{eq:SO3_CS} are \textit{coherent states} for the $\mathrm{SO}(3)$ symmetry group.

\subsubsection{Properties of the Kravchuk representation}

\begin{proposition}
\label{prop:kravchuk}
The Kravchuk transform $T$~\eqref{eq:expr_K} satisfies
\begin{enumerate}
\item $\boldsymbol{y} \rightarrow T\boldsymbol{y}$ is \textit{linear}.
\item $T$ is \textit{invertible}, with a resolution of the identity
\begin{align}
\label{eq:expr_K_inv}
\boldsymbol{y} =\frac{N+1}{4\pi} \int_{S^2} \overline{T\boldsymbol{y}(\vartheta,\varphi)} \boldsymbol{\Psi}_{\vartheta, \varphi} \, \mathrm{d}\mu(\vartheta,\varphi),
\end{align}
where  $\mathrm{d}\mu(\vartheta, \varphi) =\sin(\vartheta) \mathrm{d}\vartheta \mathrm{d}\varphi$ is the uniform measure on the sphere.
\item $T$ preserves the energy, that is,
\begin{align}
\label{eq:T_isometry}
\lVert \boldsymbol{y}\rVert_2^2 = \frac{N+1}{4\pi}  \int_{S^2} \lvert T\boldsymbol{y} (\vartheta, \varphi)\rvert^2 \, \mathrm{d}\mu(\vartheta, \varphi).
\end{align}
\item $T$ is \textit{covariant} under the action of $\mathrm{SO}(3)$, meaning that 
\begin{align}
\label{eq:T_cov}
T[\boldsymbol{R}_{\boldsymbol{u}}\boldsymbol{y}](\vartheta, \varphi) = T\boldsymbol{y}(R_{\boldsymbol{u}}(\vartheta, \varphi)),
\end{align}
where $\boldsymbol{R}_{\boldsymbol{u}}$ (resp.  $R_{\boldsymbol{u}}$) denotes the action\footnote{See Section~\ref{app:group_th} of the Supplementary material for a short presentation of the representation theory of $\mathrm{SO}(3)$.} of the rotation parameterized by the unitary vector $\boldsymbol{u} \in \mathbb{R}^3$ on vectors of size $N+1$ (resp. on points of the unit sphere). 
\item If the signal is real-valued, i.e., $\boldsymbol{y} \in \mathbb{R}^{N+1}$, then its Kravchuk spectrogram is symmetric in $\varphi$: $\forall (\vartheta, \varphi) \in [0,\pi]\times[0, 2\pi]$, $\left\lvert T\boldsymbol{y}(\vartheta, \varphi) \right\rvert^2=  \left\lvert T\boldsymbol{y}(\vartheta, 2\pi - \varphi)\right\rvert^2$.
\end{enumerate} 
\end{proposition}
\begin{proof}
Proposition~\ref{prop:kravchuk} derives from a careful translation of the properties  spin coherent states~\cite{arecchi1972atomic}, into the framework of signal processing.
For completeness,  the computations are detailed in Section~\ref{app:proof_prop} of the Supplementary material.
\end{proof}

\begin{remark}
Instead of the linear transform~\eqref{eq:expr_K}, one could follow the seminal paper \cite{atakishiyev1998wigner} and try to design a covariant Wigner-like, quadratic distribution, \textit{e.g.}, inspired by the physicists's Wigner distribution.
Yet, the theoretical study of level sets of Wigner-like distributions is intricate.
Morever, Wigner distributions usually do not come with efficient implementations.
Consequently, we focus in this paper on the Kravchuk transform and spectrogram, postponing the study of covariant discrete Wigner-like distributions like \cite{atakishiyev1998wigner} to future work.
\end{remark}

\subsection{Zeros of the Kravchuk spectrogram}
\label{ssec:zeros_spherical}

We can easily characterize the distribution of the zeros of the Kravchuk spectrogram of white Gaussian noise on $\mathbb{C}^{N+1}$.

\begin{theorem}
\label{thm:sp-gaf}
Let $\bxi \sim \mathcal{N}_\mathbb{C}(0,\boldsymbol{I})$. 
The zeros of the Kravchuk spectrogram $\lvert T\bxi(\vartheta,\varphi)\rvert^2$ of complex white Gaussian noise,
 when sent to the Riemann complex plane $\mathbb{C}\cup \lbrace \infty\rbrace$ \textit{via} the stereographic mapping 
\begin{align}
\label{eq:change_z}
( \vartheta, \varphi) \mapsto z = \cot(\vartheta/2)\mathrm{e}^{\mathrm{i}\varphi}, 
\end{align}
coincide, in law, with the zeros of the spherical Gaussian Analytic Function
\begin{align}
\label{eq:defGAFS}
\mathsf{GAF}_{\mathbb{S}}(z) = \sum_{n=0}^{N} \bxi'[n] \sqrt{\binom{N}{n}} z^n, \quad \bxi'[n] \sim\mathcal{N}_{\mathbb{C}}(0,1) \, \, \text{i.i.d.}.
\end{align}
\end{theorem}
\begin{proof}
We first rewrite the Kravchuk transform~\eqref{eq:expr_K} as a function of a complex variable, using the stereographic mapping~\eqref{eq:change_z}.
This leads to
\begin{align}
\label{eq:expr_K_lem}
T\boldsymbol{y}(z) = \frac{1}{\sqrt{(1+\lvert z \rvert^2)^N}} \sum_{n = 0}^N \sqrt{\binom{N}{n}} (\textbf{Q}\boldsymbol{y})[n] z^n,
\end{align}
where we abusively denote by $T\boldsymbol{y}$ the Kravchuk transform, either expressed as a function of the spherical coordinates $(\vartheta, \varphi)$ or of the complex stereographic variable $z$.

Now, since the Kravchuk basis 
introduced in Section~\ref{sssec:krav_basis} is orthonormal, the vector $\bxi' = \textbf{Q} \bxi$ is also a complex white Gaussian noise.
Using~\eqref{eq:expr_K_lem}, it follows that
\begin{align}
\label{e:correspondence_T_L}
T\boldsymbol{\bxi}(z) = \frac{1}{\sqrt{(1+\lvert z \rvert^2)^N}} \sum_{n = 0}^N \sqrt{\binom{N}{n}} \bxi'[n] z^n
\end{align}
is proportional to the spherical Gaussian Analytic Function defined in~\eqref{eq:defGAFS}, up to a nonvanishing prefactor.
\end{proof}

 \begin{remark}
 The rewriting of the Kravchuk transform provided in Equation~\eqref{eq:expr_K_lem} enables to connect the proposed covariant transform to the discrete transform
 $\mathscr{L}\boldsymbol{y}(z)$ of~\cite[Section 4.5]{bardenet2021time} by
 \begin{align}
 T\boldsymbol{y}(z) = \sqrt{(1+\lvert z \rvert^2)^{-N}} \times \mathscr{L}\boldsymbol{y}(z).
 \end{align}
$T$ and $\mathscr{L}$ thus differ by a non-analytic prefactor\footnote{As a side note, once we realized that the necessary prefactor was given by \eqref{e:correspondence_T_L}, we found another natural derivation of $T$ from $\mathscr{L}$; see Section~\ref{app:moments} of the supplementary material.  
Unlike the route through spin coherent states shown here, it does not easily give the covariance, though.}.
This prefactor naturally appears when defining $T$ through spin coherent states as we do in this paper, and is key in making $T$ isometric and covariant, as we showed in Proposition~\ref{prop:kravchuk}.
Finally, the prefactor also makes sure that $T$ does not explode when $\vert z\vert$ is large, which makes numerical evaluations tractable while the practical implementation of the transform $\mathscr{L}\boldsymbol{y}(z)$ of~\cite[Section 4.5]{bardenet2021time} required \textit{ad hoc} normalization\footnote{The normalization by the maximum of a well-chosen window used by~\cite{bardenet2021time} is not discussed in the paper, but can be observed in the companion {\sc Python} code at \href{https://github.com/rbardenet/tf-transforms-and-gafs/tree/452b0bbe313f5f3a32ae8dd439c07e0902ddf5c0}{https://github.com/rbardenet/tf-transforms-and-gafs}.}.
 \end{remark}

Now that we have identified the law of the zeros of the Kravchuk transform of complex white Gaussian noise, we can leverage known results on Gaussian Analytic Functions.
In particular, Theorem~\ref{thm:sp-gaf} combined with~\eqref{eq:expr_K_lem}, yields two corollaries of utmost importance in designing zero-based detection procedures in Section~\ref{sec:detect}.

\begin{corollary}~\cite[Proposition 2.3.4]{hough2009zeros}
The distribution of the zeros of the Kravchuk spectrogram of complex white Gaussian noise is invariant under the isometries of the sphere.
\end{corollary}

\begin{corollary}~\cite[Lemma 2.4.1]{hough2009zeros}
\label{cor:N_zeros}
The Kravchuk spectrogram of complex white Gaussian noise has almost surely $N$ \textit{simple} zeros.
\end{corollary}

\section{Implementation of the Kravchuk transform and extraction of the zeros}
\label{sec:implementation}

The definition~\eqref{eq:expr_K} of the Kravchuk transform has been handy to establish Theorem~\ref{thm:sp-gaf}. 
However, we explain in Section~\ref{ssec:tech_issues} why its naive implementation appears to be numerically unstable.
Therefore, in Section~\ref{ssec:gen_form}, we follow the construction of \cite{bardenet2021time} and rewrite our transform using a generating identity for Kravchuk polynomials. 
We then show that the resulting expression is amenable to computation.

\subsection{Instability of the evaluation of Kravchuk polynomials}
\label{ssec:tech_issues}

\begin{figure}
\centering
\begin{subfigure}{0.49\linewidth}
\centering
\includegraphics[width = 0.95\linewidth]{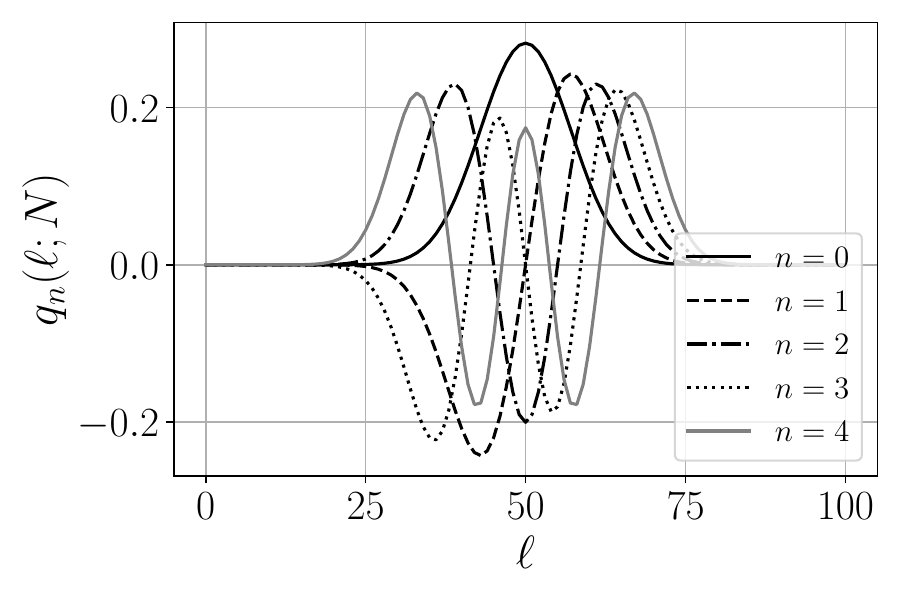}
\subcaption{Five lowest order Kravchuk functions.}
\end{subfigure}
\begin{subfigure}{0.49\linewidth}
\centering
\includegraphics[width = 0.95\linewidth]{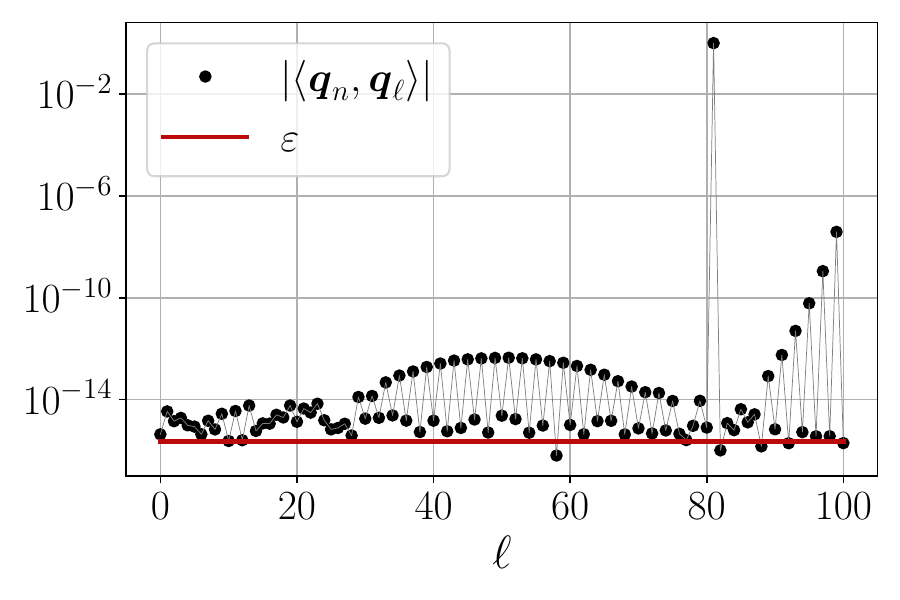}
\subcaption{\label{fig:ortho_kravchuk}Default of numerical orthogonality of Kravchuk functions.} 
\end{subfigure}
\caption{The maximal degree is $N=100$, and we consider the orthogonality of the $n = 81^{\text{th}}$ Kravchuk function with respect to the entire basis.  The bold red line at $\varepsilon = 10^{-16}$ indicates the machine precision.}
\end{figure}

The definition Equation~\eqref{eq:expr_K} of the Kravchuk transform involves the coefficients of the signal in the basis of Kravchuk functions. 
This amounts to evaluating the scalar products~\eqref{eq:decomp_Krav} for each degree $n = 0, \hdots, N$.
The most direct method to compute~\eqref{eq:decomp_Krav} requires prior evaluation at all entire points $\ell = 0, \hdots, N$ of the Kravchuk functions, themselves defined using Kravchuk polynomials~\eqref{eq:def_Krav_fun}.
In turn, the standard way to evaluate Kravchuk polynomials is to iterate the computation over the index $n$, relying on the recursion relation 
\begin{align}
\label{eq:rec_Krav}
(N-n)Q_{n+1}&(t;N) =  (N - 2t) Q_n(t;N)  - nQ_{n-1}(t;N),
\end{align}
which is provided, \textit{e.g.}, in~\cite[Chapter 6]{ismail2005classical}.
However, the coefficients involved in~\eqref{eq:rec_Krav} grow with $N$, making the recursion based on~\eqref{eq:rec_Krav} unstable as one considers signals with large number of points.
As a consequence, the practical decomposition of a signal onto the Kravchuk basis turns out to be dramatically ill-conditioned.
This is illustrated in Figure~\ref{fig:ortho_kravchuk}, where we show the lack of numerical orthogonality between the elements of the basis, even for moderate values of $n,N$. 
Without further insight, this has prevented us so far from designing a robust decomposition algorithm from the recursive evaluation of the Kravchuk polynomials.

\subsection{A stable reformulation of the Kravchuk transform}
\label{ssec:gen_form}

To obtain a stable implementation of~\eqref{eq:expr_K}, we circumvent in Proposition~\ref{prop:new_form} the problematic change from the canonical basis to the Kravchuk basis operated in Equation~\eqref{eq:decomp_Krav}.

\begin{proposition}
\label{prop:new_form}
Let $z = \cot (\vartheta/2)\mathrm{e}^{\mathrm{i} \varphi}$ denote the stereographic parameterization of Riemann's complex plane by the unit sphere. Equation~\eqref{eq:expr_K} rewrites
\begin{align}
\label{eq:rewrite_K}
T\boldsymbol{y}(z) =  \frac{1}{\sqrt{(1+\lvert z^2\rvert)^N}}\sum_{\ell = 0}^N \sqrt{\binom{N}{\ell}} \overline{\boldsymbol{y}[\ell] } \frac{\left(1-z \right)^{\ell} \left(1+z\right)^{N-\ell} }{\sqrt{2^N}}.
\end{align}
\end{proposition}
Note that \eqref{eq:rewrite_K} only involves the coefficients $\boldsymbol{y}[\ell]$ of the discrete signal $\boldsymbol{y}$ in the canonical basis of $\mathbb{C}^{N+1}$, and does not depend anymore on evaluating Kravchuk functions.

\begin{proof}
We start from a generating formula for the Kravchuk polynomials~\cite[Section~6.2]{ismail2005classical}. For all $\ell \in \lbrace 0, 1, \hdots, N\rbrace$,
\begin{align}
\sum_{n = 0}^N \binom{N}{n} Q_n(\ell;N) z^n = (1-z)^\ell(1+z)^{N-\ell}.
\end{align}
The symmetric Kravchuk functions~\eqref{eq:def_Krav_fun} thus satisfy
\begin{align}
\label{eq:gen_fun}
 \sum_{n = 0}^N \sqrt{\binom{N}{n}} q_n(\ell;N) z^n  =\sqrt{\binom{N}{\ell}} \frac{\left(1-z\right)^\ell \left(1+z\right)^{N-\ell}}{\sqrt{2^N}}.
\end{align}
On the other hand, injecting the expression of the scalar product~\eqref{eq:decomp_Krav} into the original expression of the Kravchuk transform~\eqref{eq:expr_K}, and remembering that $z = \cot(\vartheta/2)\mathrm{e}^{\mathrm{i}\varphi}$, we obtain
\begin{align}
\label{eq:K_pract}
T\boldsymbol{y}(z) = \frac{1}{\sqrt{(1+\lvert z \rvert^2)^N}} \sum_{n = 0}^N \sqrt{\binom{N}{n}} \left( \sum_{\ell = 0}^N \overline{\boldsymbol{y}[\ell]} q_n(\ell;N) \right)z^n,
\end{align}
from which we derive
\begin{align}
T\boldsymbol{y}(z) = \frac{1}{\sqrt{(1+\lvert z \rvert^2)^N}} \sum_{\ell = 0}^N\overline{\boldsymbol{y}[\ell]} \left( \sum_{n = 0}^N \sqrt{\binom{N}{n}}   q_n(\ell;N) z^n \right).
\end{align}
Finally, we rewrite the term in parentheses using~\eqref{eq:gen_fun}.
\end{proof}

\subsection{Finding zeros of Kravchuk spectrograms}
\label{ssec:MGN}

As derived at Equation~\eqref{eq:expr_K_lem}, when the Kravchuk transform is expressed as a function of the complex stereographic variable, it turns out to be proportional, up to a nonvanishing prefactor, to a polynomial of degree $N$.
Hence, extracting the zeros of the Kravchuk spectrogram amounts to finding $N$ polynomial roots.
Unfortunately, the computation of the roots of a polynomial, \textit{e.g.}, from its companion matrix, is numerically unstable for values of $N$ in the hundreds. 
For the extraction of the zeros of the Kravchuk spectrogram, we thus resort to approximate techniques.

\begin{figure}
\centering
\begin{subfigure}[t]{0.49\linewidth}
\centering
\begin{tikzpicture}[scale = 0.5]
\fill[black!75!white] (-1,-1) rectangle (1,1);

\fill[black!50!white] (-3,-3) rectangle (-1,-1);
\fill[black!50!white] (-3,1) rectangle (-1,3);
\fill[black!50!white] (1,-1) rectangle (3,1);
\fill[black!50!white] (3,3) rectangle (5,5);
\fill[black!50!white] (-1,-5) rectangle (1,-3);
\fill[black!50!white] (-5,-1) rectangle (-3,1);

\fill[black!45!white] (-5,-3) rectangle (-3,-1);
\fill[black!45!white] (-1,-3) rectangle (1,-1);
\fill[black!45!white] (1,3) rectangle (3,5);

\fill[black!40!white] (-1,1) rectangle (1,3);

\fill[black!35!white] (1,-3) rectangle (3,-1);
\fill[black!35!white] (3,1) rectangle (5,3);
\fill[black!35!white] (-5,1) rectangle (-3,3);

\fill[black!30!white] (-3,3) rectangle (-1,5);

\fill[black!25!white] (-5,-5) rectangle (-3,-3);
\fill[black!25!white] (-3,-1) rectangle (-1,1);
\fill[black!25!white] (3,-3) rectangle (5,-1);
\fill[black!25!white] (1,1) rectangle (3,3);
\fill[black!25!white] (-1,3) rectangle (1,5);

\fill[black!20!white] (3,-5) rectangle (5,-1);

\fill[black!15!white] (-3,-5) rectangle (-1,-3);
\fill[black!15!white] (1,-5) rectangle (3,-3);
\fill[black!15!white] (3,-1) rectangle (5,1);
\fill[black!15!white] (-5,3) rectangle (-3,5);

\foreach \k in {-5,-3,...,5}
	{\draw[thin,black!90!white] (\k,-5) -- (\k,5);
	\draw[thin,black!90!white] (-5,\k) -- (5,\k);}
\draw[line width = 1.15pt, color = carmin!50!white] (-1,-1) rectangle (1,1);
\draw[dashed, line width = 1.5pt, color = black] (-3,-3) rectangle (3,3);

\draw (-6,-4) node[rotate = 90]{\scriptsize $\vartheta_{j-2}$};
\draw(-6,-2) node[rotate = 90]{\scriptsize $\vartheta_{j-1}$};
\draw(-6,0) node[rotate = 90, color = carmin]{\scriptsize $\vartheta_j$};
\draw(-6,2) node[rotate = 90]{\scriptsize $\vartheta_{j+1}$};
\draw(-6,4) node[rotate = 90]{\scriptsize $\vartheta_{j+2}$};

\draw(-4,-6) node{\scriptsize $\varphi_{j-2}$};
\draw(-2,-6) node{\scriptsize $\varphi_{j-1}$};
\draw(0,-6) node[color = carmin]{\scriptsize $\varphi_j$};
\draw(2,-6) node{\scriptsize $\varphi_{j+1}$};
\draw(4,-6) node{\scriptsize $\varphi_{j+2}$};

\end{tikzpicture}
\subcaption{\label{sfig:MGN}A local minimum}
\end{subfigure}
\begin{subfigure}[t]{0.49\linewidth}
\centering
\includegraphics[width = 0.9\linewidth]{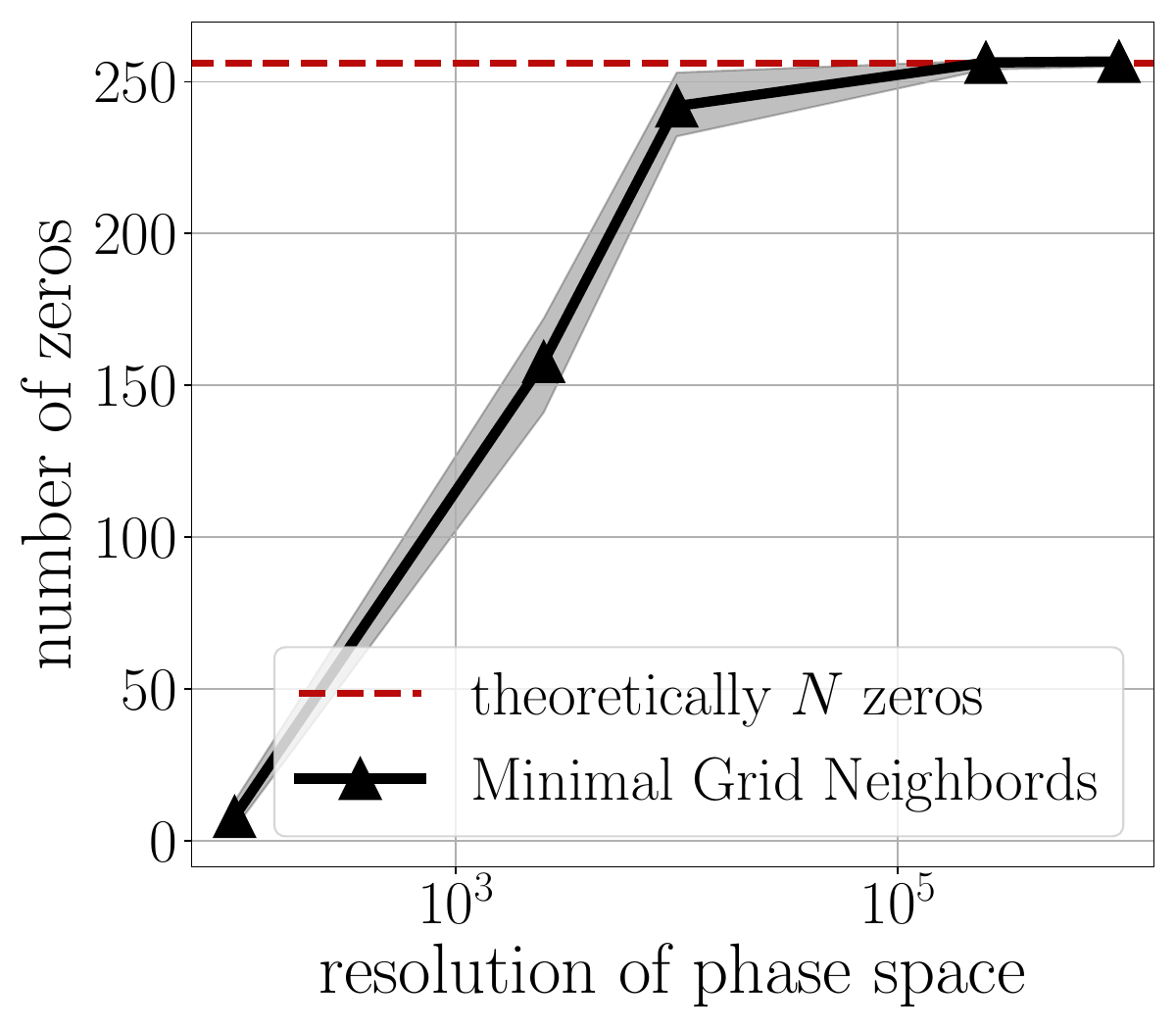}
\subcaption{\label{sfig:detected_zeros}Number of zeros found.}
\end{subfigure}
\caption{\textbf{Extraction of zeros of Kravchuk spectrograms.} A point of the phase space, red square in~(a), is considered as a spectrogram local minima as soon as the value of the spectrogram at this point is lower than all of its eight nearest neighbors, dashed black square in~(a). 
The Minimal Grid Neighbors method described in Section~\ref{ssec:MGN} is applied to 200 noisy chirps, with signal-to-noise ratio $\mathsf{snr} = 2$, for different resolution of the spherical phase space $(\vartheta,\varphi)$ and the averaged number of zeros extracted is displayed in~(b).}
\end{figure}
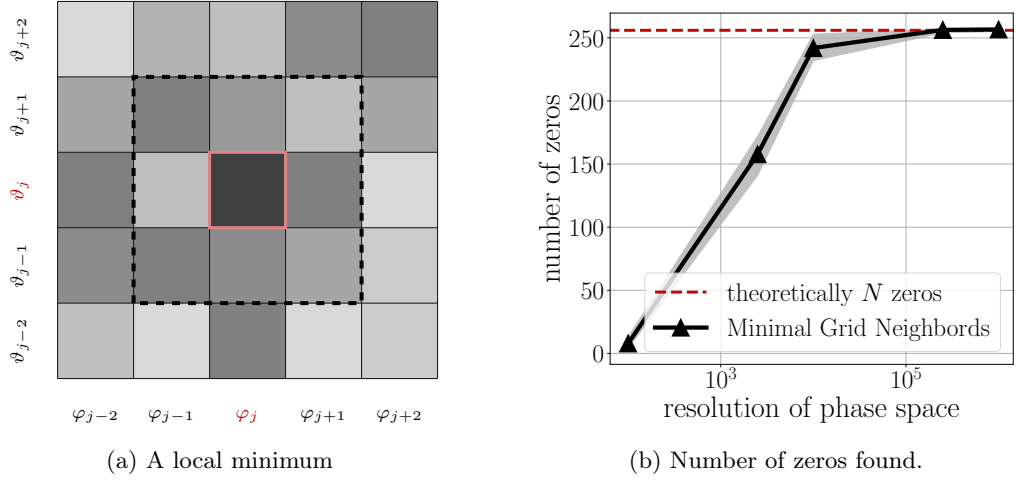

We follow the same lines as in~\cite{flandrin2015time,bardenet2020zeros}, using the method of \textit{Minimal Grid Neighbors}, illustrated at Figure~\ref{sfig:MGN}.
More precisely, assume that we have evaluated the Kravchuk spectrogram on a uniform grid on the sphere 
$$
(\vartheta,\phi)\in a\mathbb{Z}\times b\mathbb{Z} \cap [0,\pi]\times [0,2\pi],
$$ 
for some $a,b>0$.
Local minima, \textit{e.g.}, $(\vartheta_j,\varphi_j)$ in red in Figure~\ref{sfig:MGN}, are first identified as the points of the grid at which the value of the Kravchuk spectrogram is lower than the values at its eight nearest neighbors, represented by the bold dashed square in Figure~\ref{sfig:MGN}.
Then, all the local minima inferior to a pre-specified threshold are considered as numerical zeros.
To the best of our knowledge, a similar method is used in all practical studies involving the zeros of Fourier spectrograms~\cite{flandrin2015time,bardenet2020zeros,bardenet2021time} and~\cite[Chapters 13 and 15]{flandrin2018explorations}, or scalogram zeros~\cite{abreu2018filtering}, although using a threshold might not be necessary in the Fourier case \cite[Theorem~1]{abreu2020local}. 

Compared to the case of Fourier spectrograms discussed in Section~\ref{ssec:zeros_tf},  the main advantage of the Kravchuk spectrogram is that, thanks to~Corollary~\ref{cor:N_zeros}, we know that in the white noise case, it has almost surely $N$ simple zeros.
Furthermore, the zeros arise from the noise structure, thus it is reasonable to expect that, as soon as the noise level is moderate, the same proposition applies to Kravchuk spectrogram of noisy signals.
This enables a very simple assessment of the accuracy of the extracted set of zeros, and it circumvents technical considerations to compare the number of extracted zeros to their expected number in~\cite{bardenet2020zeros}.
In particular, the threshold used in the extraction of zeros can be chosen by checking that this condition is fulfilled.
In practice, we observed that a threshold of $7.5\%$ of the maximum amplitude of the Kravchuk spectrogram is perfectly adequate for a large range of both the signal-to-noise ratio $\mathsf{snr}$ and number $N$ of points in the input signal.
Another key setting of the Minimal Grid Neighbors method is the resolution of the grid on which the spectrogram is computed.
We plot on Figure~\ref{sfig:detected_zeros} the number of zeros detected for different resolutions of the grid, averaged over 200 realizations.
As soon as the resolution $N_{\vartheta} \times N_{\varphi}$ is large enough,  the expected $N$ zeros are indeed detected, up to intrinsic randomness, which validates the Minimal Grid Neighbors approach.

\subsection{Kravchuk representation of noisy chirps}

\begin{figure*}
\centering
\begin{subfigure}{0.19\linewidth}
\centering
\includegraphics[width = \linewidth]{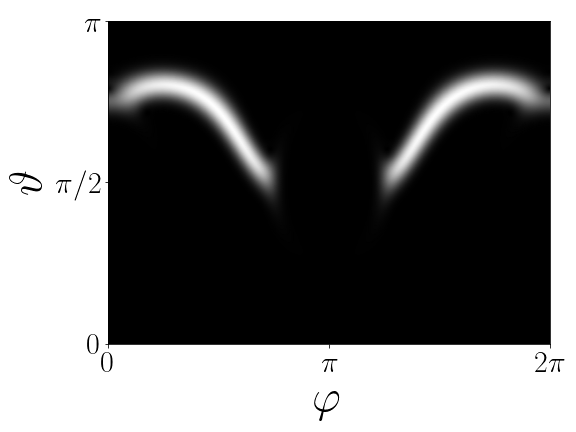}\\

\vspace{2mm}
\includegraphics[width = 0.75\linewidth]{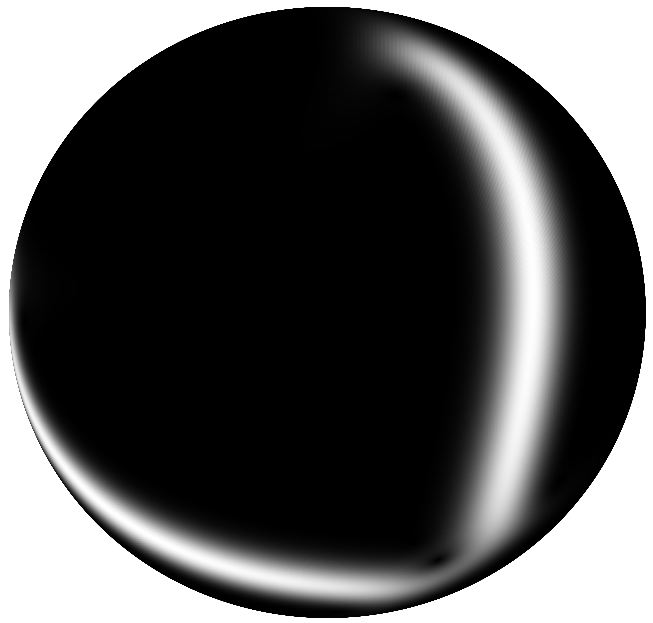}
\subcaption{$\mathsf{snr} = \infty$}
\end{subfigure}
\begin{subfigure}{0.19\linewidth}
\centering
\includegraphics[width = \linewidth]{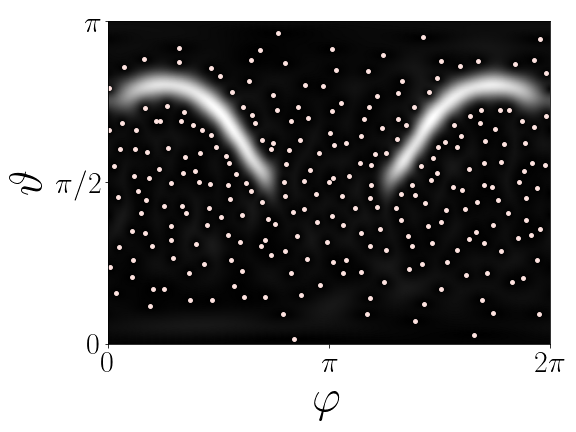}\\

\vspace{2mm}
\includegraphics[width = 0.75\linewidth]{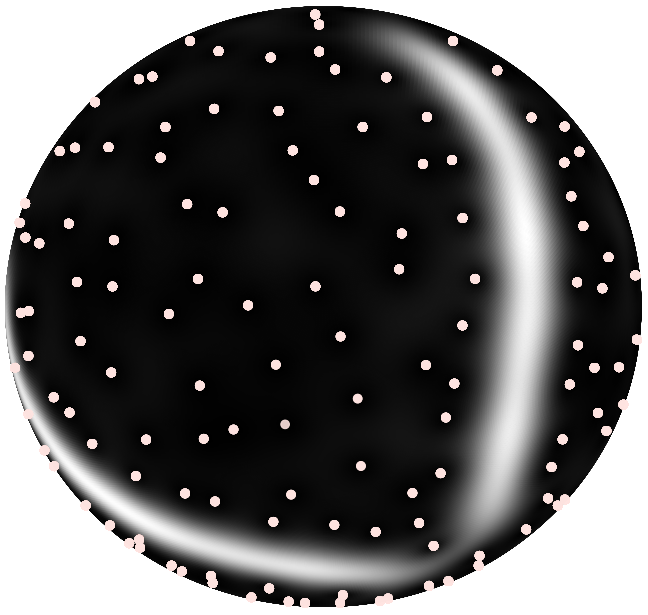}
\subcaption{$\mathsf{snr} = 5$}
\end{subfigure}
\begin{subfigure}{0.19\linewidth}
\centering
\includegraphics[width = \linewidth]{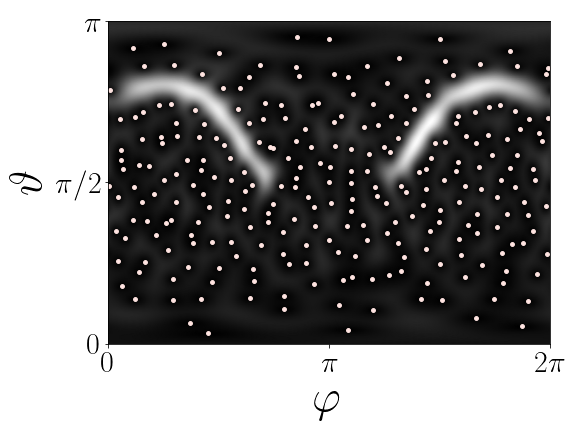}\\

\vspace{2mm}
\includegraphics[width = 0.75\linewidth]{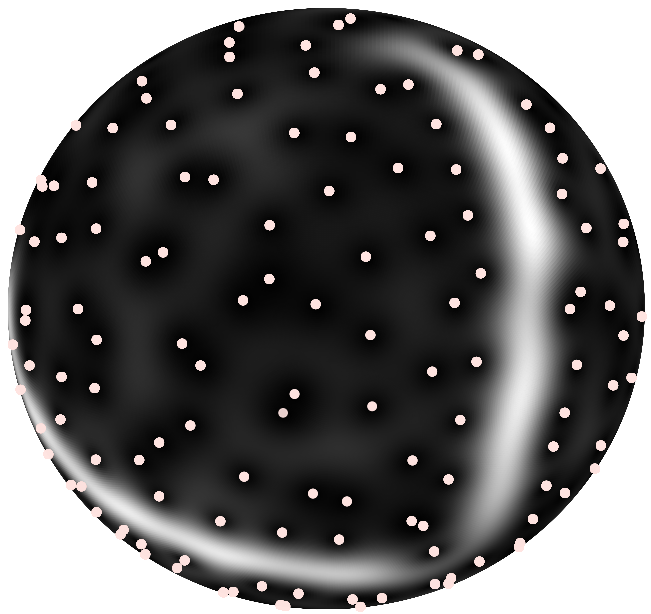}
\subcaption{$\mathsf{snr} = 2$}
\end{subfigure}
\begin{subfigure}{0.19\linewidth}
\centering
\includegraphics[width = \linewidth]{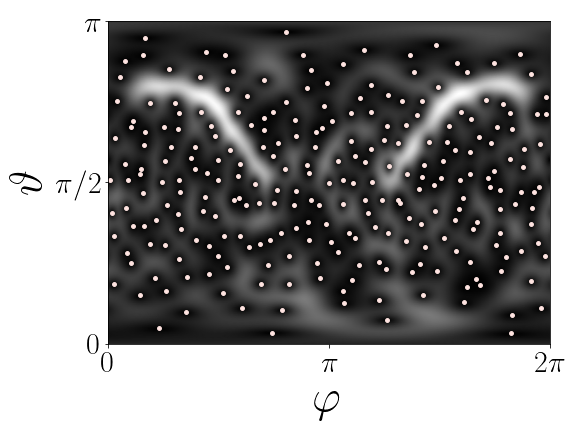}\\

\vspace{2mm}
\includegraphics[width = 0.75\linewidth]{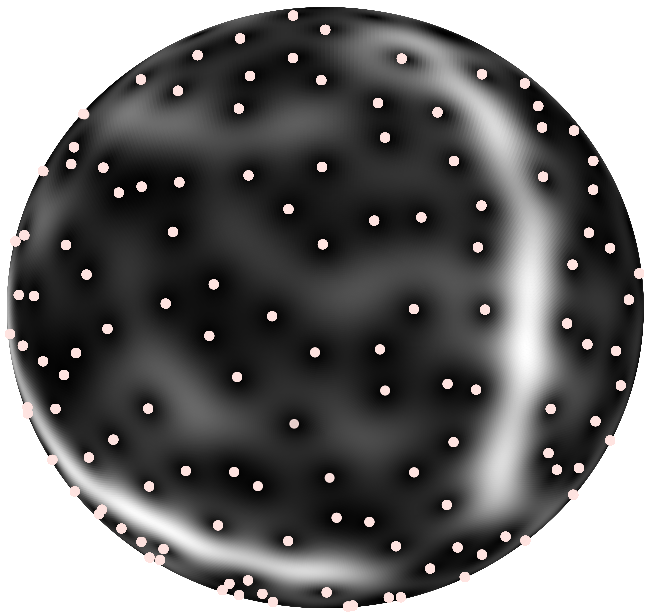}
\subcaption{$\mathsf{snr} = 1$}
\end{subfigure}
\begin{subfigure}{0.19\linewidth}
\centering
\includegraphics[width = \linewidth]{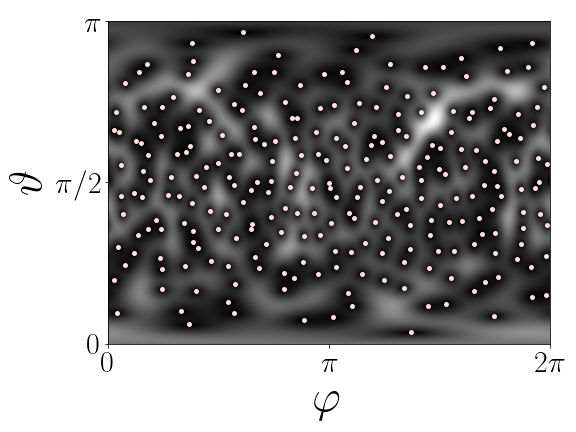}\\

\vspace{2mm}
\includegraphics[width = 0.75\linewidth]{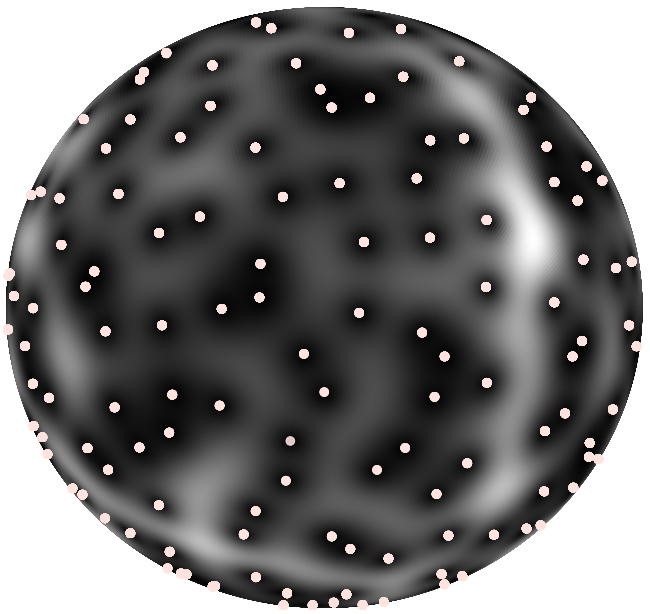}
\subcaption{$\mathsf{snr} = 0.5$}
\end{subfigure}
\caption{\label{fig:nkravchuk}\textbf{Kravchuk spectrogram of noisy chirps.} For each of the signals of Figure~\ref{fig:nchirp},  the squared modulus of the proposed Kravchuk transform~\eqref{eq:expr_K} is displayed in grey level as a function of the spherical coordinates $(\vartheta, \varphi)$, in an unfolded representation (top row) and in the natural spherical geometry (bottom row), with zeros indicated by pale rose dots. The signal-to-noise ratio is decreasing from left to right.}
\end{figure*}

Direct implementation of Formula~\eqref{eq:rewrite_K} permits to compute the Kravchuk transform of the noisy chirps signals of Figure~\ref{fig:nchirp}, the squared modulus of which yields the associated Kravchuk specrogram, then, the Minimal Grid Neighbors method described at Section~\ref{ssec:MGN} provides the zeros,  altogether leading to Figure~\ref{fig:nkravchuk}.
First, a planar representation in $(\vartheta, \varphi)$ coordinates is provided in Figure~\ref{fig:nkravchuk}, top row, on which the symmetry in $\varphi$ for real signals can be clearly observed; see Proposition~\ref{prop:kravchuk},~5).
Second, a direct representation on the sphere, bottom row of Figure~\ref{fig:nkravchuk}, illustrates the uniform spread of the zeros, outside of the phase space region corresponding to the signal.
This very regular behavior of the zeros in the absence of signal illustrates Theorem~\ref{thm:sp-gaf}, as the zeros of the spherical GAF are known to be a repulsive point process \cite{hough2009zeros}. 
Furthermore, the fact that the signal repels the spectrogram zeros is in perfect agreement with previous observations in the Fourier case~\cite{gardner2006sparse,flandrin2015time, bardenet2020zeros}, and is at the core of the development of signal processing procedures based on the zeros of the spectrogram.

\section{A detection procedure based on the zeros of the Kravchuk spectrogram}
\label{sec:detect}

As observed in Figure~\ref{fig:nkravchuk}, the presence of a signal induces local perturbations in the pattern formed by the zeros of the Kravchuk spectrogram: \textit{holes} appears in the distribution of zeros in the regions of the phase space corresponding to the signal.
Consequently, the random configuration of zeros deviates from the regularly spread point process of Figure~\ref{sfig:GAFS} obtained for pure noise.
In this section, we follow the same lines as for the classical Short-Time Fourier transform \cite{bardenet2020zeros} and turn 
Theorem~\ref{thm:sp-gaf} into nonparametric statistical tests for detecting the presence of some signal embedded into white noise.

\subsection{General principle of hypothesis testing}

 We aim at discriminating between the null hypothesis $\textbf{H}_0$, ``the observations consist in pure noise", and the alternative $\textbf{H}_1$, ``the data contain a deterministic signal of interest".
Mathematically, we consider the two situations
\begin{align*}
\textbf{H}_0: \quad \boldsymbol{y} = \boldsymbol{\xi}, \quad \quad \textbf{H}_1: \quad \boldsymbol{y} =\mathsf{snr} \times \boldsymbol{x} + \boldsymbol{\xi}
\end{align*}
where $\boldsymbol{\xi}$ denotes the complex white Gaussian noise and $\boldsymbol{x}$ is an unknown deterministic waveform, \textit{e.g.}, a sampled chirp
of the form~\eqref{eq:def_chirp}, and $\mathsf{snr} > 0$ is the signal-to-noise ratio.

To design a detection procedure, we use a summary statistic $s(\boldsymbol{y})\in\mathbb{R}$, such that measuring large value of $s$ advocates for rejecting the null hypothesis.
For the test to be efficient, $s$ should quantify precisely the discrepancy between the data and pure noise.

We consider Monte Carlo tests, characterized by a level of significance $\alpha$, a number of samples under the null hypothesis $m$ and an index $k$, chosen so that $\alpha = k/(m+1)$.
Once these parameters are fixed, testing data $\boldsymbol{y}$ consists in going through the following steps: \textit{(i)}~generate $m$ independent samples of complex white Gaussian noise and compute their summary statistics $s_1\geq s_2\geq \hdots\geq s_m$ sorted in decreasing order; \textit{(ii)}~compute the summary statistics of the observations $\boldsymbol{y}$ under concern; \textit{(iii)}~if $s(\boldsymbol{y}) \geq s_k$, then reject the null hypothesis with confidence $1-\alpha$.

A key point in constructing detection tests based on the zeros of the Kravchuk spectrogram lies in the design of appropriate summary statistics $s$, enabling to discriminate between the pure noise situation in which zeros are evenly spread on the sphere, such as in Figure~\ref{sfig:GAFS}, and signal plus noise cases, in which holes appears in the zeros pattern, as in the Kravchuk spectrograms in Figure~\ref{fig:nkravchuk}.
To that aim, we turn to the toolbox of spatial statistics , specifically developed for the analysis of \textit{point processes}, \emph{i.e}, random point patterns.

\subsection{Spatial statistics on zeros of spectrogram}
\label{ssec:geosto}

\subsubsection{Point processes}

Theorem~\ref{thm:sp-gaf} and Equation~\eqref{eq:expr_K_lem} insures that the zeros of the Kravchuk transform of a noisy signal are almost surely $N$ isolated points lying on the unit sphere. 
In particular, the set of zeros is a \textit{point process} on the sphere equipped with the \textit{chordal} distance
\begin{align}
\label{eq:def_chord}
  \mathsf{d}&\left((\vartheta_1,\varphi_1),(\vartheta_2,\varphi_2)\right)= \arccos \left( \sin \vartheta_1 \sin \vartheta_2 \cos(\varphi_1-\varphi_2) + \cos \vartheta_1 \cos \vartheta_2\right).
\end{align}
Formally a point process $Z$ is a distribution over configurations of points in a metric space,  characterized by its \textit{spatial statistics}.
The simplest, first order, spatial statistics is the \textit{density} $\rho : S^2 \rightarrow \mathbb{R}_+$ satisfying, if it exists, 
\begin{align}
\label{eq:density}
\forall U \subseteq S^2, \quad \mathbb{E} \left[ \mathsf{card}( Z \cap U )\right] = \int_{U} \rho(\vartheta,\varphi) \, \mathrm{d}\mu (\vartheta, \varphi),
\end{align}
where $\mu$ is the uniform measure on the sphere defined in Proposition~\ref{prop:kravchuk}, and $\mathsf{card}$ denotes the cardinality of a set, so that the left-hand side of~\eqref{eq:density} counts the expected number of points of the point process falling into $U$.

If the point process is invariant under isometries of $S^2$, \textit{e.g.}, if $Z$ consists in the zeros of the spherical Gaussian Analytic Function displayed in Figure~\ref{sfig:GAFS}, it is said to be \textit{stationary}, and its density is constant.
The interest reader is referred to~\cite{moller2003statistical} for further definitions and properties.

\subsubsection{Functional statistics}

As illustrated on Figure~\ref{fig:nkravchuk}, the presence of some signal creates some holes in the zeros pattern.
The presence of such holes, modifies the distribution of distances between zeros,
advocating for the use of \textit{second order} spatial statistics to discriminate between the signal plus noise and the pure noise cases.
We will consider two of them,  benefiting from robust estimators which can be implemented efficiently.

First, \textit{Ripley's $K$ function} accounts for the distribution of the pair distances, and is proportional, for each $r>0$, to the expected number of pairs at distance less than $r$~\cite{ripley1976second}.
The standard definition initially proposed by~\cite[Chapter 4]{moller2003statistical} for point processes in $\mathbb{R}^d$,  has very recently been adapted to the case of stationary point processes on $S^2$~\cite[Section 3.2]{moller2016functional} defining 
\begin{align}
\label{eq:def_K}
K(r) = \frac{1}{4\pi \rho^2} \mathbb{E} \sum_{(\vartheta_i,\varphi_i) \in Z}^{\neq}   \boldsymbol{1} \left(\mathsf{d}\left((\vartheta_1,\varphi_1),(\vartheta_2,\varphi_2)\right) < r \right)
\end{align}
where the sum runs over all pairs $\left((\vartheta_1,\varphi_1),(\vartheta_2,\varphi_2)\right)$ of distinct points in $Z$.
where $\rho$ denotes the constant density of the point process $Z$, $4\pi$ is the surface of the unit sphere and $\boldsymbol{1}$ denotes the indicator of an event, taking value one or zero depending on whether the condition is fulfilled.

Second, the \textit{empty space function} $F$ of a \textit{stationary} point process is the distribution function of the distance from the origin, or equivalently to any fixed point of the space due to the stationarity of the point process, to the nearest point of the point process.
Direct adaptation of the definition proposed by~\cite{moller2003statistical}, lead to the definition:
\begin{align}
\label{eq:def_F}
F(r) = \mathbb{P}\left(\mathsf{b}(0,r) \cap Z\neq \emptyset \right),
\end{align}
where $\mathbb{P}$ is the probability measure over the realizations of the point process $Z$, and $\mathsf{b}(0,r)$ denotes the ball centered at the origin and of radius $r > 0$ for the chordal distance.

\subsubsection{Practical estimators}
\label{ssec:sum_stat_est}

Performance of the testing procedure rely on the ability to estimate accurately the functional statistics, which encodes the characteristics of the point process made of the zeros of the Kravchuk spectrogram.
We review shortly nonparametric estimators for each of the two functional statistics $K$ and $F$. 
Further considerations are discussed in~\cite{moller2007modern}.

Ripley's $K$ function being linked to the pair distances,  estimating $K(r)$ amounts to count the number of pairs of zeros which are at chordal distance less than $r$.
Then,  for a \textit{stationary} point process $Z$ on the sphere
\begin{align}
\label{eq:hat_k}
\widehat{K}(r) = \frac{(4\pi)^2}{N_z} \sum_{(\vartheta_i,\varphi_i) \in Z}^{\neq} \boldsymbol{1}\left( \mathsf{d} \left( (\vartheta_1,\varphi_1),(\vartheta_2,\varphi_2)\right)\leq r\right), 
\end{align}
where $N_Z$ is the empirical number of points in $Z$, yields an unbiased estimator of Ripley's $K$ function.

\begin{remark}
Note that, thanks to Corollary~\ref{cor:N_zeros}, we know that $N_Z = N$ almost surely.
Thus $N_Z$ could be replaced by $N$ in~\eqref{eq:hat_k}, reducing the variance the estimate.
In practice, the empirical number of zeros differs from $N$ by at most one and we observed no difference in the result of the detection.
\end{remark}

The empty space function $F$ accounts for the distribution of the size of the holes in the zeros pattern.
Let $\lbrace (\vartheta_j, \varphi_j), \, j = 1, \hdots, N_{\#} \rbrace$ a uniform grid on the sphere.
The practical estimation of $F$ requires to count how many points of the grid lie at distance less than $r$ from a point of $Z$.
An unbiased estimate of the empty space function of a stationary point process $Z$ on the sphere is thus given by
\begin{align}
\label{eq:hat_F}
\widehat{F}(r) =\frac{1}{N_{\#}} \sum_{j = 1}^{N_{\#}} \boldsymbol{1}\left(\inf_{(\vartheta, \varphi) \in Z}  \mathsf{d}\left((\vartheta_j,\varphi_j),(\vartheta,\varphi)\right)  < r\right).
\end{align}

\subsection{Monte Carlo envelope testing}
\label{ssec:monte_carlo}

The methodology of envelope testing~\cite{baddeley2014tests} being the same whatever the chosen function statistics, we describe it for a generic functional $S(r)$, which should be thought of as either Ripley's $K$ function~\eqref{eq:def_K} or the empty space function~\eqref{eq:def_F}.
A relevant summary statistics $s$ should measure precisely the discrepancy between the functional statistics estimated from the data, $\widehat{S}_{\boldsymbol{y}}(r)$, to the reference functional statistics of the zeros of the Kravchuk spectrogram of complex white Gaussian noise, $S_0(r)$.
To that aim, following~\cite{bardenet2020zeros}, we construct the summary statistics
\begin{align}
\label{eq:test_stats}
s(\boldsymbol{y}) = \sqrt{\int_{r_1}^{r_2} \left\lvert \widehat{S}_{\boldsymbol{y}}(r) - S_0(r) \right\rvert^2 \, \mathrm{d}r},
\end{align}
which quantifies the quadratic distance between the estimated functional statistics and the expected functional under the null hypothesis.

Though, to the best of our knowledge, nor the Ripley's $K_0$ function,  neither the empty space function $F_0$ of the point process of the zeros of the spherical Gaussian Analytic Function, corresponding the the pure noise reference case, have a documented explicit expression.
In practice, the theoretical functional statistic $S_0(r)$ involved in the definition of the summary statistics~\eqref{eq:test_stats}, is hence replaced by an empirical averaged $\bar{S}_0(r)$ over the functional statistics estimated from the $m$ realizations of complex white Gaussian noise and from the data
\begin{align}
\label{eq:barS}
\bar{S}_0 = \frac{1}{m+1} \left(\widehat{S}_1 + \hdots + \widehat{S}_m + \widehat{S}_{\boldsymbol{y}}\right).
\end{align}
Interestingly, it has been demonstrated in~\cite{baddeley2014tests} that replacing the theoretical functional statistics by the pointwise average~\eqref{eq:barS} does not impair the significance of the Monte Carlo envelope test.

\section{Experiments}
\label{sec:exp}

The detection procedure based on the zeros of the Kravchuk specrogram designed in Section~\ref{sec:detect} is assessed on synthetic data.
The performance of the test is investigated, varying the characteristics of the signal, and hence the difficulty of the task.
Furthermore, we compare the proposed strategy to the state-of-the-art detection procedure based on the zeros of the Fourier spectrogram.

\subsection{Settings}
\subsubsection{Synthetic data}

Numerical simulations focus on the detection of deterministic chirps following the parametric Model~\eqref{eq:def_chirp}, corrupted by a superimposed complex white Gaussian noise, according to~\eqref{eq:def_data}. 
The discrete signals considered consist in these noisy chirps, sampled at $N+1$ points, regularly spaced in a temporal window of $40$~s.
The characteristic frequencies of the chirps are fixed at $f_1 = 0.5 $~Hz and $f_2 = 1.25$~Hz, while the duration of the chirp $\nu$ and the length of the observation $N$ are varied.
The noise level is controlled by the signal-to-noise ratio $\mathsf{snr}$, introduced at Equation~\eqref{eq:def_data}.
In practice, both the deterministic signal $ \boldsymbol{x}$ and the additive noise $ \boldsymbol{\xi}$ are $\ell^2$-normalized, \textit{i.e.}, $\lVert \boldsymbol{x} \rVert_2 = \lVert \boldsymbol{\xi} \rVert_2$, so that the noise level only depends on $\mathsf{snr}$, and not on the characteristics of the chirp.  
Example of noisy chirps of duration~$2\nu = 30$~s with $N= 513$ points and decreasing signal-to-noise ratios are provided in Figure~\ref{fig:nchirp}.

\subsubsection{Estimation of functional statistics}

Ripley's $K$ function is estimated using the unbiased estimator provided at Equation~\eqref{eq:hat_k},  for $10^4$ points ranging from $r_1 = 0$ to $r_2 = \pi$, the maximal possible distance on the sphere, as can be seen from~\eqref{eq:def_chord}.
As for the estimation of the empty space function $F$, we use of the estimator~\eqref{eq:hat_F}, with a grid $(\vartheta_i,\varphi_i)$ of resolution $N_{\#} = 4\sqrt{N_Z} \times 4\sqrt{N_Z}$, where $N_Z$ denotes the empirical number of zeros of the Kravchuk spectrogram.
$\widehat{F}(r)$ is computed at $10^4$ points,  for $r$ ranging from $r_1 = 0$ to $r_2 = 2\pi/\sqrt{N}$, as we observed that $\widehat{F}(r)$ was saturating at 1 for larger values of $r$.

\subsubsection{Power assessment} 
The Monte Carlo testing methodology designed in Section~\ref{sec:detect} is run with systematic significance level $\alpha = 0.05$, relying on $m = 199$ noise realizations, and hence corresponding to comparing the observed summary statistic to the $k = 10^{\text{th}}$ largest value obtained under the null hypothesis.
To measure the performance of the designed detection test for given duration $\nu$, number of points $N$, and signal-to-noise ratio $\mathsf{snr}$, 200 independent noisy chirps are generated from the observation model~\eqref{eq:def_data}.
Then the test is run, choosing either the $K$ or the $F$ functional statistic, and the averaged number of detection yields the estimated power of the test $\widehat{\beta}$.
The quality of $\widehat{\beta}$ as an estimator for the power of the test is assessed using Clopper-Pearson confidence intervals~\cite{brown2001interval} at level $0.01$. 
We chose this value for ease of mental computation: for an experiment summarized with 10 intervals, for instance, a simple Bonferroni correction~\cite{wasserman2004all} thus allows jointly considering all intervals at significance level $0.1$. 

\subsection{Detection performance}
\label{ssec:detection_perf}

\subsubsection{Choice of the functional statistics}

\begin{figure}
\centering
\begin{subfigure}{0.49\linewidth}
\centering
\includegraphics[width = \linewidth]{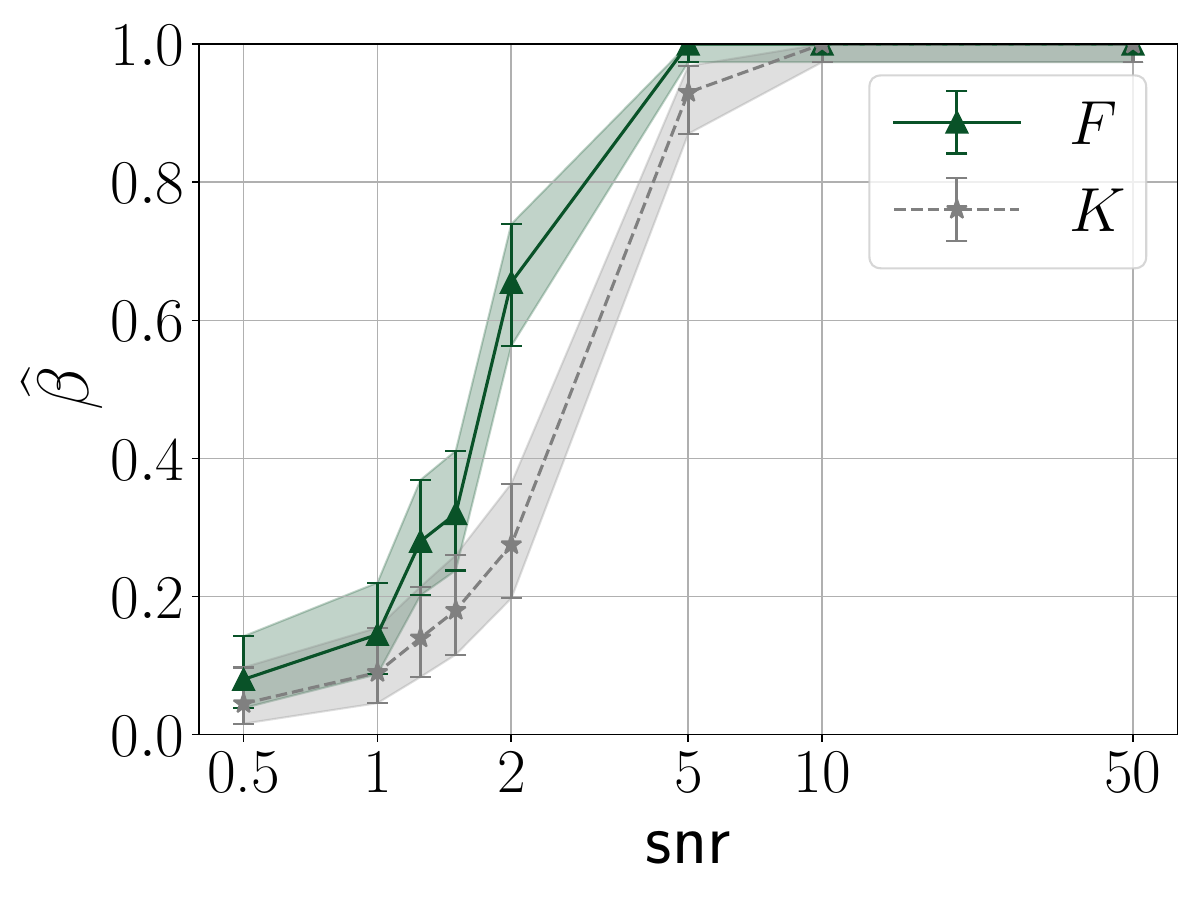}
\subcaption{$N+1 = 257$ points}
\end{subfigure}
\begin{subfigure}{0.49\linewidth}
\centering
\includegraphics[width = \linewidth]{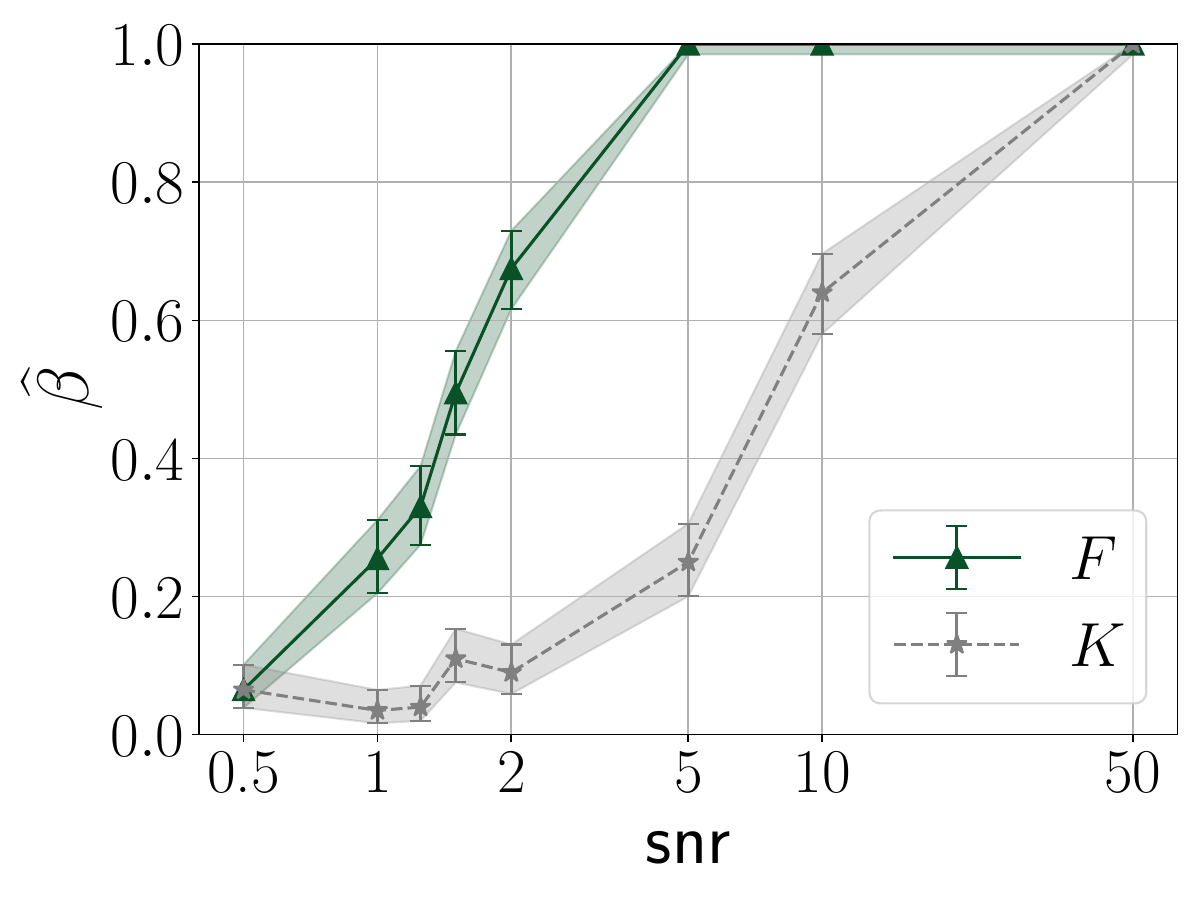}
\subcaption{$N+1 = 513$ points}
\end{subfigure}
\caption{\label{fig:K_or_F} \textbf{Comparison between $K$ and $F$ functional statistics.}
Evolution of the power of the test with the signal-to-noise ratio.}
\end{figure}

We consider noisy chirps of duration $2\nu = 30$~s, with $N+1 \in \lbrace 257, 513\rbrace$ sample points, for eight different signal-to-noise ratio $\mathsf{snr} \in \lbrace 0.5, 1, 1.25, 1.5, 2, 5, 10, 50\rbrace$ and compare the power of the detection test based on the zeros of their Kravchuk spectrogram when using either Ripley's $K$ function or the empty space function $F$ for defining the summary statistic $s(\boldsymbol{y})$.
First, as expected, we observe in Figure~\ref{fig:K_or_F} that the power of the test increases as the signal-to-noise ratio increases, \textit{i.e.}, the easier the detection, the better the test performance.
Second, whatever the noise level $\mathsf{snr}$, the test using the empty space function $F$ has a significantly higher power than the one using Ripley's $K$ function.
Similar conclusions were obtained for other signal lengths and chirp durations.
Since, like for Fourier spectrograms \cite{bardenet2020zeros}, the empty space function systematically yields larger power for the same significance, we henceforth focus on the empty space function.

\subsubsection{Influence of the characteristics of the signals}

Intuitively, the detection task is all the more difficult that: \textit{(i)}~the signal-to-noise ratio is low, \textit{(ii)}~the duration of the chirp is small compared to the length of the observation window, and \textit{(iii)}~the number  of sampling points $N+1$ is small.
In order to verify these statements, we run systematic tests on signals of different lengths $N\in \lbrace 128, 256, 512, 1024 \rbrace$, two different durations $2\nu \in \lbrace 20~\text{s}, 30~\text{s} \rbrace$ for a fixed observation window of $40$~s, and different signal-to-noise ratio $\mathsf{snr} \in \lbrace 1.5, 2\rbrace$.

\begin{figure}
\centering
\begin{subfigure}{0.49\linewidth}
\centering
\includegraphics[width = \linewidth]{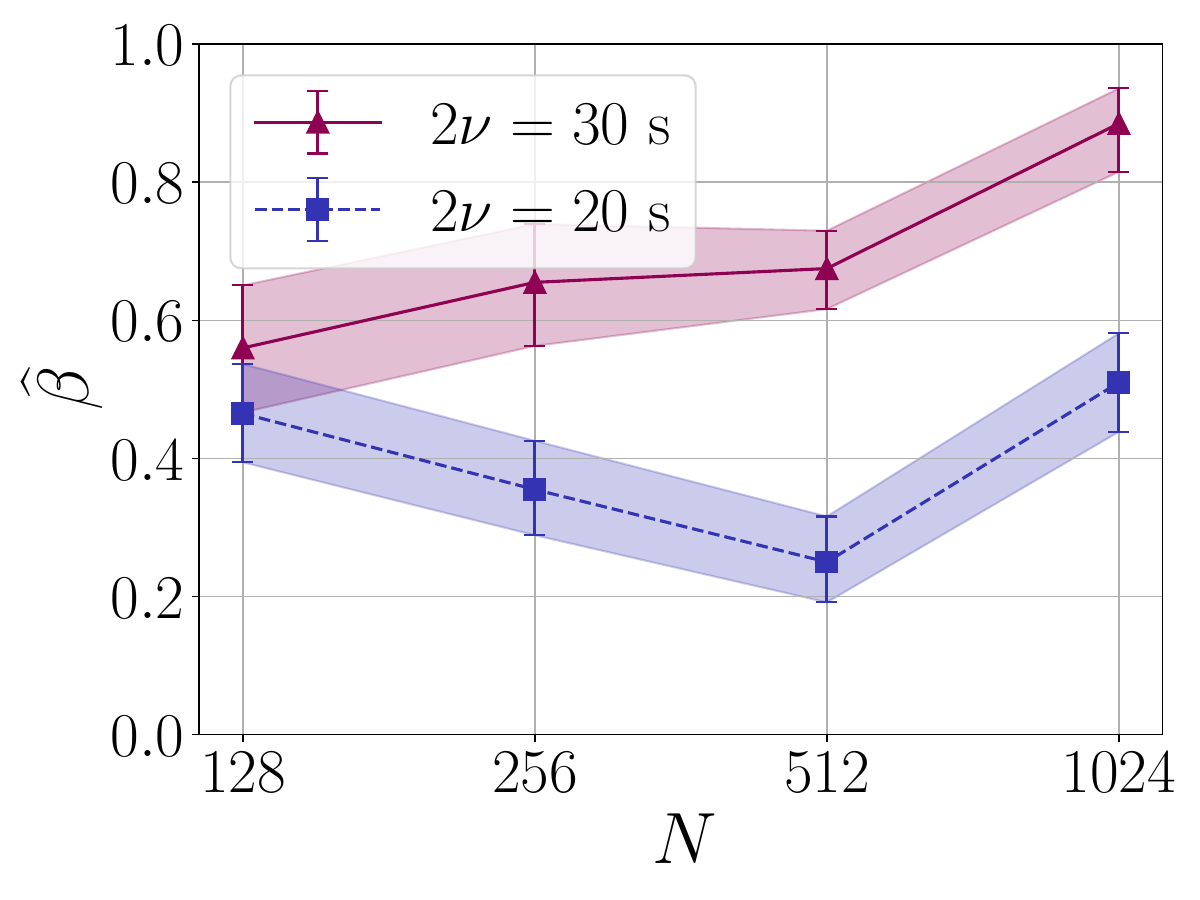}
\subcaption{\label{sfig:power_N_duration_2}$\mathsf{snr} = 2$}
\end{subfigure}
\begin{subfigure}{0.49\linewidth}
\centering
\includegraphics[width = \linewidth]{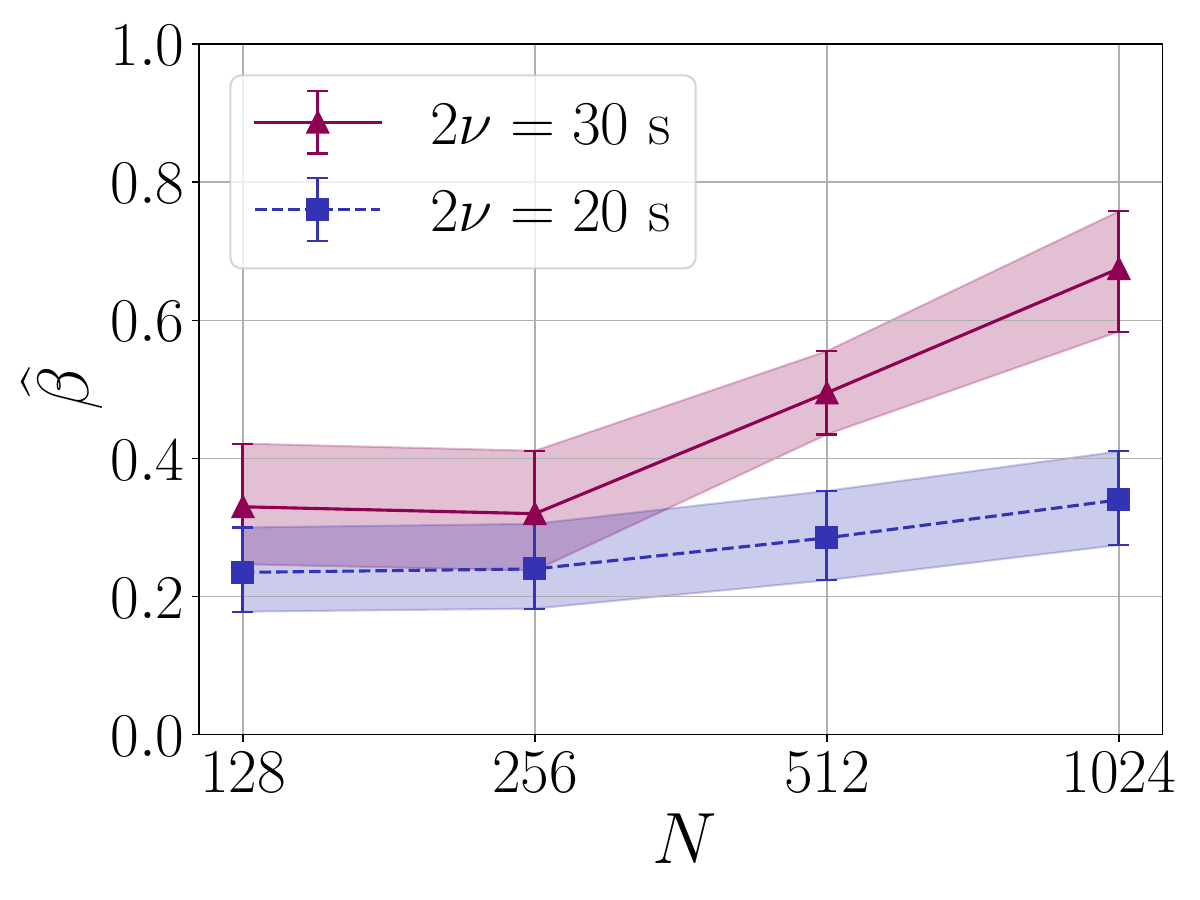}
\subcaption{$\mathsf{snr} = 1.5$}
\end{subfigure}
\caption{\label{fig:nu_N} \textbf{Robustness to small number of samples and short duration.}
Evolution of the power of the test with the length of the observation.}
\end{figure}

In the easier configuration, $\mathsf{snr} = 2$ and $2\nu =30$,  in magenta on~Figure~\ref{sfig:power_N_duration_2},  increasing the number of point increases the power of the test.
As the detection problem gets harder, either because of lower signal-to-noise ratio, or shorter duration,  increasing the number of points is not enough to improve performance.
This observation led us to conjecture that the number of points is not a critical feature and that, as soon as $N$ is large enough, functional statistics are accurately estimated, and the detection procedure is only limited by the difficulty of the task.
This could indicate that the proposed methodology possesses a regime in which it is independent of the sampling rate, which turns very interesting for processing real-world signals.
Furthermore, the magenta solid curve, corresponding to signal of larger duration is always above the blue dashed curve, assessing that the power of the test is larger when the chirp is longer.

\subsubsection{Kravchuk vs. Fourier spectrograms}

We now compare to the zero-based detection test proposed by~\cite{bardenet2020zeros}, relying on the zeros of the standard Fourier spectrogram described in Section~\ref{sec:stateofart}.
Tests are run on noisy chirps with fixed signal-to-noise ratio $\mathsf{snr} = 1.5$, and we explore the robustness of power against the number of sampled points in both an easy situation, corresponding to chirps of duration $2\nu = 30$~s, and a difficult one, corresponding to $2\nu = 20$~s.

\begin{figure}
\centering
\begin{subfigure}{0.49\linewidth}
\centering
\includegraphics[width = \linewidth]{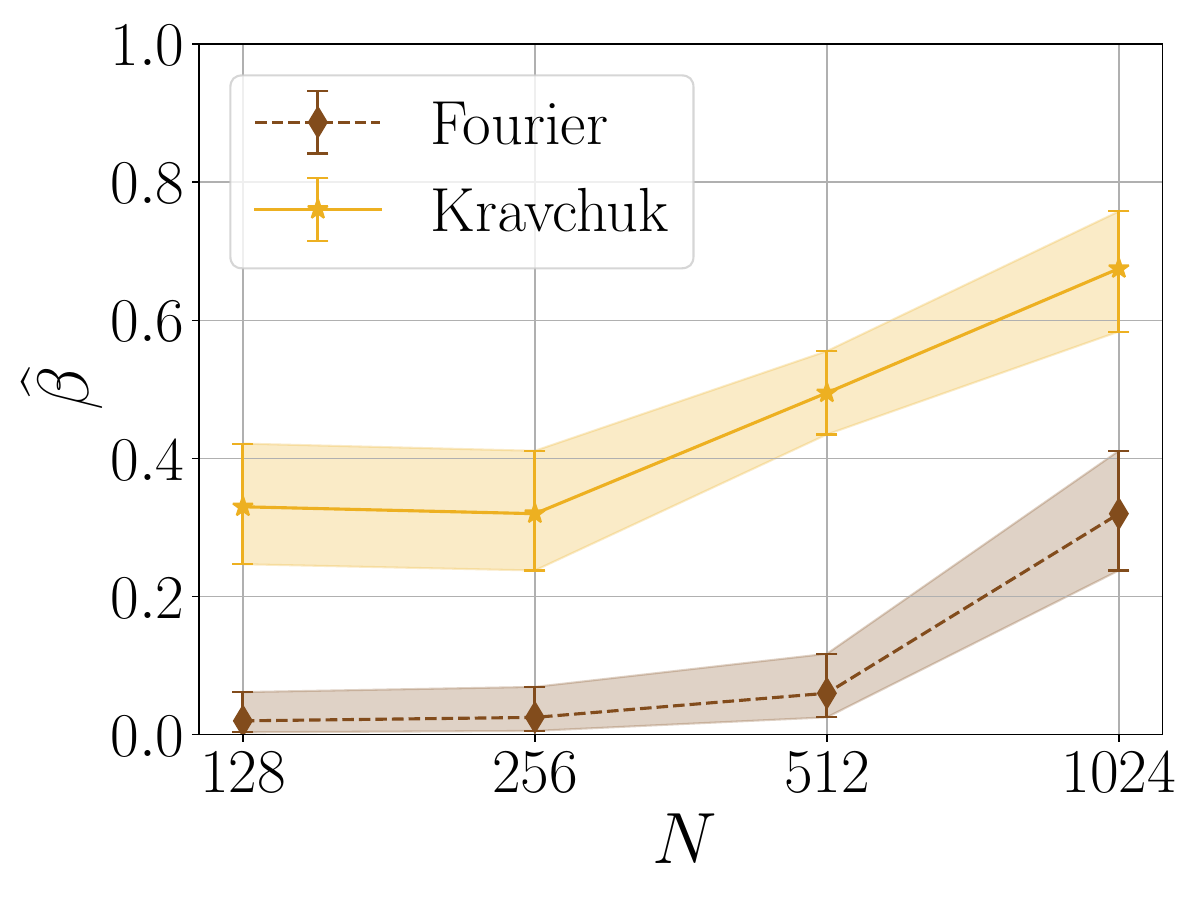}
\subcaption{duration $ 2\nu =30$~s}
\end{subfigure}
\begin{subfigure}{0.49\linewidth}
\centering
\includegraphics[width = \linewidth]{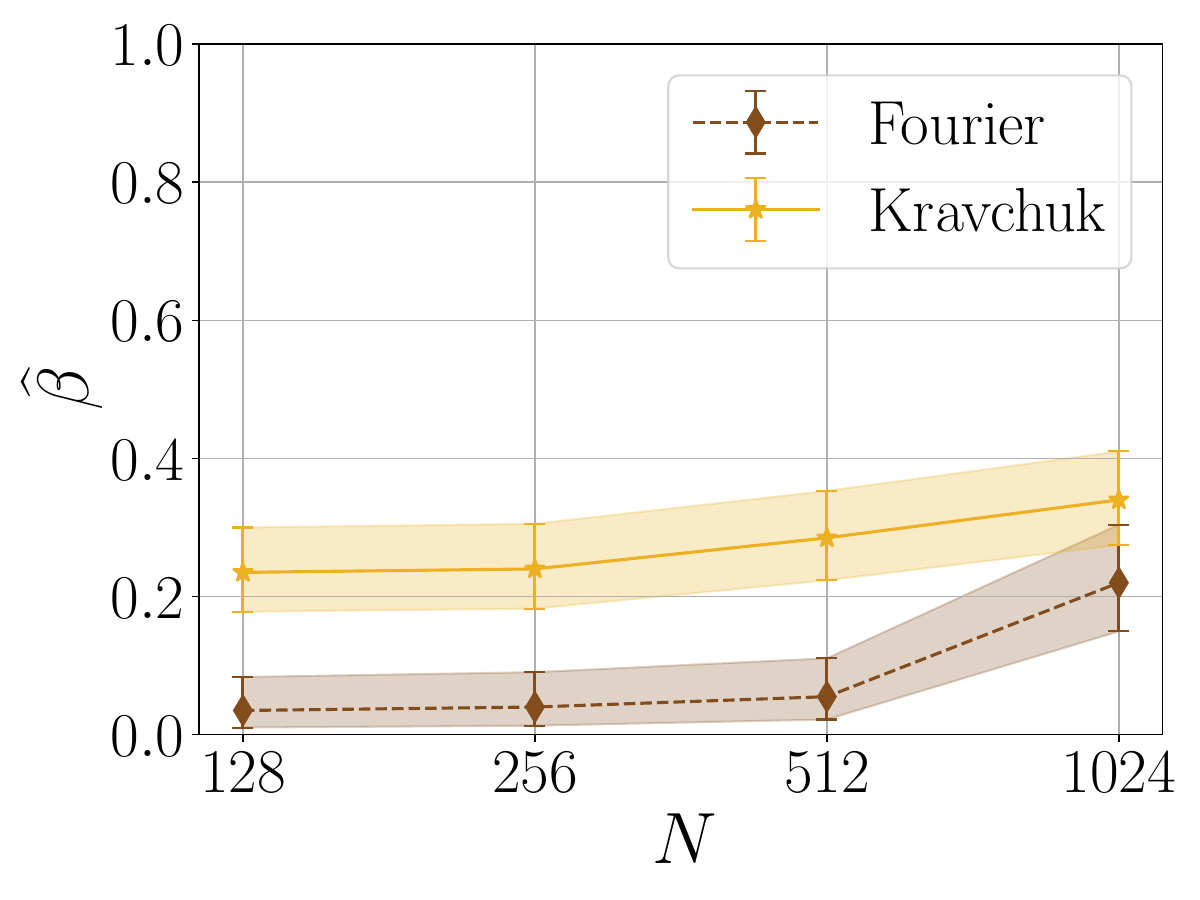}
\subcaption{\label{sfig:Fourier_or_Kravchuk_10} duration $ 2\nu =20$~s}
\end{subfigure}
\caption{\label{fig:Fourier_or_Kravchuk} \textbf{Detection tests based on the zeros of either Fourier or Kravchuk spectrogram.}
Evolution of the power of the test with the signal length $N$ for noisy signals with fixed $\mathsf{snr} = 1.5$.}
\end{figure}

We observe in Figure~\ref{fig:Fourier_or_Kravchuk} that the Kravchuk-based detection, corresponding to the yellow solid line, systematically outperforms the Fourier-based detection, corresponding to the brown dashed line.
Furthermore, the power of the test decreases more slowly as $N$ decreases in the case of Kravchuk spectrogram, especially for chirps of shorter duration, as shown in Figure~\ref{sfig:Fourier_or_Kravchuk_10}.

The better performance of the detection strategy based on Kravchuk spectrograms can be explained by two core properties of the Kravchuk transform.
First, it has been specifically designed for discrete signals, hence its computation is exact and does not induce information loss.
Second, the phase space associated with the Kravchuk representation is compact, consequently the entire point process of zeros is observed and the estimation of functional statistics is direct, not requiring sophisticated edge corrections.
In other words, the characteristic patterns reflecting the presence of a signal are more faithfully rendered by the Kravchuk representation than by the traditional discrete approximation to the Short-Time Fourier transform with Gaussian window. 
These patterns are then more precisely captured by functional statistics on the sphere, which is compact, compared to the unbounded time-frequency plane.

\begin{remark}
The signal detection experiments based on the zeros of the Fourier spectrogram in~\cite[Section~5.2]{bardenet2020zeros} were performed on signals normalized in amplitude, contrary to the $\ell^2$ normalization used in the present work.
Consequently the signal-to-noise ratios cannot be compared. 
In particular, the detection problems considered in our Section~\ref{ssec:detection_perf} are more difficult than those tackled in~\cite{bardenet2020zeros}, explaining the poor performance observed in Figure~\ref{fig:Fourier_or_Kravchuk} when using the Fourier spectrogram.
\end{remark}

\section{Conclusions}

Motivated by the desire to find a time-frequency interpretation of seminal transforms based on Kravchuk polynomials introduced in \cite{bardenet2021time}, and by analogies with spin coherent states in quantum optics,
we introduced a new covariant representation, the Kravchuk spectrogram, tailored for discrete signals.
The phase space is the unit sphere, and we showed that the zeros of the  Kravchuk spectrogram of complex white Gaussian noise have the same distribution as the zeros of the spherical Gaussian Analytic Function. 
In particular, the zeros are invariant under isometries of the sphere.
Leveraging the stationarity of the zeros, we demonstrated that Monte Carlo envelope tests based on spherical functional statistics yield powerful detection tests. 
Compared to Fourier spectrograms \cite{bardenet2020zeros}, the Kravchuk representation bypasses both the need to discretize the continuous Short-Time Fourier Transform, and the need for edge correction of functional statistics estimators. 
Intensive numerical simulations demonstrate that these advantages lead to more powerful detection tests than Fourier spectrograms, in particular when the signal-to-noise ratio or the number of samples is low.
Another advantage of our procedure is its nonparametric aspect, along with the absence of hyperparameters.

We now list a few avenues for future work.
While our implementation circumvents the instability of evaluating Kravchuk polynomials, it requires $\mathcal O(N^3)$ operations for each point of the grid we put on the phase space.
While this is enough for small signals, say $N\lesssim 1024$, a fast implementation of the Kravchuk transform would significantly broaden its applicability.
We are currently working on a fast scheme, consisting in a rotation-covariant counterpart of the Fast Fourier Transform algorithm.
Then, taking advantage of the reconstruction formula~\eqref{eq:expr_K_inv} and of the compactness of the phase space,  we will construct new zero-based denoising and AM-FM component separation algorithms. 
We expect the latter to outperform previous procedures in some regimes of practical interest~\cite{flandrin2015time}, notably when the Riemann approximations to the continuous Fourier transforms involved in the classical Short-Time Fourier transform are inaccurate.
Furthermore, we plan to adapt the recent extraction of zeros of~\cite{escudero2021efficient}, which comes with more theoretical guarantees than the Minimal Grid Neighbors approach.
Again, the advantage of the Kravchuk transform here is that we can evaluate it pointwise, unlike the classical Short-Time Fourier transform which needs to be approximated.

On the stochastic geometry side, spatial statistics on the sphere have received interest lately~\cite{moller2016functional,moller2018determinantal,hirao2021finite}. 
For starters, the Kravchuk transform applied to white noise can be seen as a way to approximately sample the zeros of the spherical Gaussian Analytic Function. 
Moreover, real-valued noisy signals yield a \emph{symmetric} spherical Gaussian Analytic Function, in the spirit of the symmetric Gaussian Analytic Functions of \cite{Fel13}. 
Results on the zeros of this random polynomial with real coefficients are bound to have an interest in signal processing. 

Finally, a large body of work in mathematical quantum physics, including contraction theorems~\cite{ricci1986contraction}, suggest that, as the number of sample points grows, the Kravchuk transform converges in some sense to an Short-Time Fourier transform with Gaussian window. 
A rigorous statement on this convergence could mean that the Kravchuk transform is a natural way of discretizing the continuous Short-Time Fourier transform while preserving a covariant structure.

A {\sc Python} toolbox is publicly available on the GitHub page of the first author\footnote{\url{https://github.com/bpascal-fr/kravchuk-transform-and-its-zeros}}, enabling to reproduce all the experiments presented in Section~\ref{sec:exp}.

\appendix

\section{Elements of group theory for $\mathrm{SO}(3)$}
\label{app:group_th}

For the sake of completeness, we summarize the main tools of group theory that are needed to understand the covariance of the Kravchuk transform in Proposition~\ref{prop:kravchuk}~4).
A detailed and rigorous presentation can be found in~\cite[Chapter 6]{kosmann2010groups}.

Essentially, we are reviewing below the construction of a particular covariant family of signals, called \textit{spin coherent states} \cite{arecchi1972atomic,gazeau2009coherent}. 
To build such a family, we need a group and a unitary representation of that group that acts on signals. 
Here the group shall be $\mathrm{SO}(3)$ and the space of signals $\mathbb{C}^{N+1}$.
This is a counterpart of the construction of the Short-Time Fourier transform, where the Weyl-Heisenberg group is used a phase space for signals in $L^2(\mathbb{R})$; see Section~\ref{ssec:algebra_CS}.

\subsection{A geometrical description of $\mathrm{SO}(3)$}

The group of rotations $\mathrm{SO}(3)$ acts on vectors of $\mathbb{R}^3$, preserving the Euclidean norm. 
In particular, $\mathrm{SO}(3)$ acts transitively on the unit sphere $S^2$ of $\mathbb{R}^3$.
The sphere is the phase space of the Kravchuk transform, and $\mathrm{SO}(3)$ is to that phase space what the Weyl-Heisenberg group of time-frequency shifts is to the time-frequency plane; see Section~\ref{ssec:algebra_CS}.

The group $\mathrm{SO}(3)$ can be parameterized by the sphere $S^2$, seen as a collection of unit vectors
\begin{align}
\boldsymbol{u}(\vartheta,\varphi) = \left(\sin(\vartheta) \cos(\varphi), \sin(\vartheta) \sin(\varphi),\cos (\vartheta)\right).
\end{align}
More precisely, the rotation of angle $\vartheta$, with axis directed by the unit vector of cartesian coordinates $(-\sin(\varphi), \cos(\varphi),0)$, is denoted by $R_{\boldsymbol{u}}$.
Note that, for the sake of clarity, we shall omit the dependency of $\boldsymbol u$ to $(\vartheta,\varphi)$ when not explicitly needed.

By construction, the rotation $R_{\boldsymbol{u}}$ sends the north pole of the sphere, of spherical coordinates $(0,0)$, onto the point of spherical coordinates $(\vartheta,\varphi)$, 
\begin{align*}
R_{\boldsymbol{u}(\vartheta,\varphi)}(0,0) = (\vartheta,\varphi).
\end{align*}
Moreover, the successive application of two rotations $R_{\boldsymbol{u}}$ and $R_{\boldsymbol{u}'}$ is still a rotation.
Hence it is associated to a unit vector, denoted by $\boldsymbol{u}\cdot \boldsymbol{u}'$. 
The group law of $\mathrm{SO}(3)$ is encapsulated into the product $\cdot$, namely
\begin{align}
R_{\boldsymbol{u}\cdot \boldsymbol{u}'} = R_{\boldsymbol{u}} \circ R_{\boldsymbol{u}'}.
\label{eq:group_law}
\end{align}
A precise description of the composition law $\boldsymbol{u}\cdot \boldsymbol{u}'$ can be found in~\cite[Section 6.5]{gazeau2009coherent}.

\subsection{A specific finite-dimensional representation}
\label{ssec:spinJ}

A linear \textit{representation} of the symmetry group $\mathrm{SO}(3)$ on a vector space $\mathcal{H}$ is an application 
\begin{align*}
\mathcal{R} : \left\lbrace
\begin{array}{ccc}
\mathrm{SO}(3) & \rightarrow & \mathrm{GL}(\mathcal{H}) \\
R_{\boldsymbol{u}} & \mapsto & \boldsymbol{R}_{\boldsymbol{u}}
\end{array}
\right.
\end{align*}
where $\mathrm{GL}(\mathcal{H})$ denotes the linear group of $\mathcal{H}$, which preserves the group law, \textit{i.e.}, satisfies
\begin{align}
\label{eq:lin_rep}
\forall \boldsymbol{u}, \boldsymbol{u}' \in S^2, \,  \boldsymbol{R}_{\boldsymbol{u}\cdot \boldsymbol{u}'}  = \boldsymbol{R}_{\boldsymbol{u}} \boldsymbol{R}_{\boldsymbol{u}'}.
\end{align}
Note that the product in the right-hand side of~\eqref{eq:lin_rep} is the product of linear operators of $\mathcal{H}$.
The representation $\boldsymbol{R}$ transposes the natural action of $\mathrm{SO}(3)$ on vectors of $S^2$ to an action on vectors of $\mathcal{H}$, with the same composition structure.

For each fixed  $N \in \mathbb{N}$, we henceforth focus on the \textit{unitary} 
representation of $\mathrm{SO}(3)$
on $\mathcal{H}=\mathbb{C}^{N+1}$ described in~\cite[Section 6.5]{gazeau2009coherent}, which is 
at the core of the quantum theory of spin-$J$, for $J = N/2$~\cite{arecchi1972atomic,gazeau2009coherent}.
Unitarity of the representation amounts to impose that the operators $\boldsymbol{R}_{\boldsymbol{u}}$ are unitary with respect to the Hermitian inner product of $\mathbb{C}^{N+1}$.
This is a central property ensuring covariance of the family of coherent states~\cite[Chapter 5]{gazeau2009coherent}.

\subsection{Action on coherent states}

The action of $\mathrm{SO}(3)$ on vectors of $\mathbb{C}^{N+1}$ described by the spin-$J$ representation of Section~\ref{ssec:spinJ}
is in general abstract, stemming from algebraic rules for the construction of group representations~\cite{kosmann2010groups}.
Yet, there exists a family of vectors on which this action is more transparent, namely the family of $\mathrm{SO}(3)$ coherent states.

Indeed, the family of coherent states defined in Equation~\eqref{eq:SO3_CS} can be obtained by letting the rotations act on what is called a \textit{mother wavelet} $\boldsymbol{\Psi}_{(0,0)}$~\cite[Section 6.5]{gazeau2009coherent}, \emph{i.e.},
\begin{align}
\boldsymbol{\Psi}_{(\vartheta,\varphi)} = \boldsymbol{R}_{\boldsymbol{u}(\vartheta,\varphi)} \boldsymbol{\Psi}_{(0,0)},
\label{eq:mother_wavelet}
\end{align}
where $\boldsymbol{\Psi}_{(0,0)} = \boldsymbol{q}_N\in\mathbb{C}^{N+1}$ is the Kravchuk function of highest degree in dimension $N+1$.

More generally, combining \eqref{eq:group_law} and \eqref{eq:mother_wavelet} shows that $\mathrm{SO}(3)$ acts in a covariant way on the family of coherent states.
Indeed, considering the action of rotation $R_{\boldsymbol{u}}$ on the point of spherical coordinates $(\vartheta',\varphi')$. 
The conservation \eqref{eq:group_law} of the group structure ensures that, in $S^2$,  $\boldsymbol{u}(\vartheta,\varphi) \cdot \boldsymbol{u}'(\vartheta',\varphi') = R_{\boldsymbol{u}'}(\vartheta,\varphi)$.
By \eqref{eq:mother_wavelet}, the action on coherent states thus writes in the covariant form
\begin{align}
\label{eq:act_CS}
\boldsymbol{R}_{\boldsymbol{u}(\vartheta,\varphi)}^* \boldsymbol{\Psi}_{(\vartheta',\varphi')}  = \boldsymbol{\Psi}_{R_{\boldsymbol{u}}(\vartheta',\varphi')}
\end{align}
where $\boldsymbol{R}_{\boldsymbol{u}}^*$ denotes the adjoint operator of $\boldsymbol{R}_{\boldsymbol{u}}$.
Finally, note that, anticipating the invertibility of the coherent state decomposition provided in Proposition~\ref{prop:kravchuk},~2) and proved in Section~\ref{app:proof_prop}, the action of $\mathrm{SO}(3)$ on any vector of $\mathbb{C}^{N+1}$ can be derived by considering its decomposition onto the family of coherent states.

\section{Proof of Proposition \ref{prop:kravchuk}}
\label{app:proof_prop}

Proposition~\ref{prop:kravchuk} is obtained by adapting the properties of spin coherent states, a tool initially introduced in quantum mechanics~\cite{arecchi1972atomic}.
We refer to \cite[Chapter 6]{gazeau2009coherent} for a modern account on spin coherent states.

\begin{proof}
\begin{enumerate}
\item The computation of the coefficients of $\boldsymbol{y}$ in the Kravchuk basis, 
\begin{align*}
\textbf{Q} : \left\lbrace \begin{array}{ccc}
\mathbb{C}^{N+1} & \rightarrow & \mathbb{C}^{N+1} \\
\boldsymbol{y} & \mapsto & \textbf{Q}\boldsymbol{y} 
\end{array}\right.
\end{align*}
is linear. 
The Kravchuk transform~\eqref{eq:expr_K} is a linear combination of the Kravchuk coefficients $(\textbf{Q}\boldsymbol{y})[n]$, and is thus linear.
\item In order to prove the reconstruction formula and the energy conservation, we use the following lemma.
\begin{lemma}{\cite[Section 6.3]{gazeau2009coherent}}
\label{lem:ortho}
As $S^2$ is equipped with the uniform measure $\mathrm{d}\mu(\vartheta,\varphi) = \sin(\vartheta) \, \mathrm{d}\vartheta\mathrm{d}\varphi$, for all $N \in \mathbb{N}^*$, the family of complex-valued functions defined on the unit sphere
\begin{align}
\psi_n(\vartheta,\varphi) =  \sqrt{\frac{N+1}{4\pi}} \sqrt{\binom{N}{n}} \left( \cos \frac{\vartheta}{2} \right)^n \left( \sin \frac{\vartheta}{2} \right)^{N-n } \mathrm{e}^{\mathrm{i}n \varphi}
\end{align}
is orthonormal in $L^2(S^2)$.
\end{lemma}

Let $\ell \in \lbrace 0, \hdots, N\rbrace$.
We show that the $\ell^{\text{th}}$ component of the right-hand side of~\eqref{eq:expr_K_inv} equals $\boldsymbol{y}[\ell]$.
Using the coherent state interpretation of the Kravchuk transform,
\begin{align}
\label{eq:right-hand}
\frac{N+1}{4\pi} \int_{S^2} \overline{T\boldsymbol{y}(\vartheta,\varphi)} \boldsymbol{\Psi}_{\vartheta, \varphi}[\ell] \, \mathrm{d}\mu(\vartheta,\varphi)
 =& \frac{N+1}{4\pi}\int_{S^2} \overline{\langle\boldsymbol{y}, \boldsymbol{\Psi}_{\vartheta,\varphi}\rangle} \boldsymbol{\Psi}_{\vartheta, \varphi}[\ell] \, \mathrm{d}\mu(\vartheta,\varphi) \nonumber\\
  =& \frac{N+1}{4\pi} \int_{S^2} \sum_{n = 0}^N \boldsymbol{y}[n] \overline{\boldsymbol{\Psi}_{\vartheta,\varphi}[n]} \boldsymbol{\Psi}_{\vartheta, \varphi}[\ell] \, \mathrm{d}\mu(\vartheta,\varphi)
\end{align}
Then, one can remark that, by definition of $\mathrm{SO}(3)$ coherent states~\eqref{eq:SO3_CS}, using the family $\lbrace \psi_n, \, n = 0,1, \hdots, N\rbrace$, they alternatively rewrite as
$
\boldsymbol{\Psi}_{\vartheta,\varphi} = \sqrt{4\pi/(N+1)} \sum_{n = 0}^N \psi_n(\vartheta,\varphi) \boldsymbol{q}_n, 
$ 
and hence, 
\begin{align}
\label{eq:prod_Psi}
 \overline{\boldsymbol{\Psi}_{\vartheta,\varphi}[n]} \boldsymbol{\Psi}_{\vartheta, \varphi}[\ell] 
&\nonumber=\frac{4\pi}{N+1} \left( \sum_{m=  0}^N \overline{\psi_m(\vartheta,\varphi)} \boldsymbol{q}_m[n]\right) \left( \sum_{m'=  0}^N \psi_{m'}(\vartheta,\varphi) \boldsymbol{q}_{m'}[\ell]\right)\\
&= \frac{4\pi}{N+1} \sum_{m=  0}^N \sum_{m'=  0}^N \overline{\psi_m(\vartheta,\varphi)}   \psi_{m'}(\vartheta,\varphi) \boldsymbol{q}_m[n] \boldsymbol{q}_{m'}[\ell].
\end{align}
Then, using Lemma~\ref{lem:ortho}, 
\begin{align}
\label{eq:int_prod_Psi}
\int_{S^2} \, \overline{\boldsymbol{\Psi}_{\vartheta,\varphi}[n]} \boldsymbol{\Psi}_{\vartheta, \varphi}[\ell] \mathrm{d}\mu(\vartheta,\varphi) 
& =\frac{4\pi}{N+1}  \sum_{m=  0}^N \sum_{m'=  0}^N \delta_{m,m'}\boldsymbol{q}_m[n] \boldsymbol{q}_{m'}[\ell]\nonumber\\
& = \frac{4\pi}{N+1} \sum_{m=  0}^N  \boldsymbol{q}_m[n] \boldsymbol{q}_{m}[\ell].
\end{align}
Injecting the above result in~\eqref{eq:right-hand} yields
\begin{align*}
 \frac{N+1}{4\pi}\int_{S^2} \overline{T\boldsymbol{y}(\vartheta,\varphi)} \boldsymbol{\Psi}_{\vartheta, \varphi}[\ell] \, \mathrm{d}\mu(\vartheta,\varphi)
& = \sum_{n = 0}^N \boldsymbol{y}[n] \left( \sum_{m=  0}^N  \boldsymbol{q}_m[n] \boldsymbol{q}_{m}[\ell] \right)\\
& =  \sum_{m=  0}^N \left(\sum_{n = 0}^N \boldsymbol{y}[n] \boldsymbol{q}_m[n] \right) \boldsymbol{q}_{m}[\ell]\\
& = \sum_{m = 0}^N \overline{ \textbf{Q}\boldsymbol{y}[m] }\boldsymbol{q}_m[\ell]\\
& = \boldsymbol{y}[\ell].
\end{align*}
\item The energy in the phase space writes
\begin{align*}
\frac{N+1}{4\pi} \int_{S^2} \lvert T\boldsymbol{y} (\vartheta, \varphi) \rvert^2 \, \mathrm{d}\mu(\vartheta, \varphi)
&=\frac{N+1}{4\pi} \int_{S^2} \lvert \langle \boldsymbol{y}, \boldsymbol{\Psi}_{\vartheta, \varphi} \rangle \rvert^2 \, \mathrm{d}\mu(\vartheta, \varphi)\\
&=\frac{N+1}{4\pi} \int_{S^2} \sum_{n=0}^N \sum_{n'=0}^N \overline{ \boldsymbol{y}[n]}  \boldsymbol{\Psi}_{\vartheta, \varphi} [n] \boldsymbol{y}[n'] \overline{\boldsymbol{\Psi}_{\vartheta, \varphi} [n']} \, \mathrm{d}\mu(\vartheta, \varphi).
\end{align*}
Then, making use of~\eqref{eq:prod_Psi} and \eqref{eq:int_prod_Psi}, we get
\begin{align*}
\frac{N+1}{4\pi} \int_{S^2} \lvert T\boldsymbol{y} (\vartheta, \varphi) \rvert^2 \, \mathrm{d}\mu(\vartheta, \varphi)
& =  \sum_{n}^N \sum_{n'}^N \overline{ \boldsymbol{y}[n]}\boldsymbol{y}[n'] \sum_{m=  0}^N  \boldsymbol{q}_m[n] \boldsymbol{q}_{m}[n']\\
& =  \sum_{m=  0}^N \left( \sum_{n}^N  \overline{ \boldsymbol{y}[n]}   \boldsymbol{q}_m[n]\right) \left( \sum_{n'}^N \boldsymbol{y}[n']  \boldsymbol{q}_{m}[n']\right)\\
& =  \sum_{m=  0}^N \textbf{Q}\boldsymbol{y}[m] \overline{\textbf{Q}\boldsymbol{y}[m]}\\
& = \lVert \textbf{Q}\boldsymbol{y} \rVert_2^2 = \lvert \boldsymbol{y}\rVert_2^2
\end{align*}
by orthonormality of the Kravchuk basis.
\item The covariance of the Kravchuk transform stems from its interpretation as a coherent state decomposition. 
Indeed, using the unitarity of the representation of $\mathrm{SO}(3)$,
\begin{align}
T[\boldsymbol{R}_{\boldsymbol{u}}\boldsymbol{y}](\vartheta, \varphi) &= \langle \boldsymbol{R}_{\boldsymbol{u}} \boldsymbol{y}, \boldsymbol{\Psi}_{(\vartheta,\varphi)} \rangle\\
&= \langle \boldsymbol{y}, \boldsymbol{R}_{\boldsymbol{u}}^* \boldsymbol{\Psi}_{(\vartheta,\varphi)} \rangle.
\end{align}
Then, from the action~\eqref{eq:act_CS} of $\mathrm{SO}(3)$ on the family of coherent states,
\begin{align*}
\langle \boldsymbol{y}, \boldsymbol{R}_{\boldsymbol{u}}^* \boldsymbol{\Psi}_{(\vartheta,\varphi)} \rangle = \langle \boldsymbol{y},  \boldsymbol{\Psi}_{R_{\boldsymbol{u}}(\vartheta,\varphi)} \rangle.
\end{align*}
Finally, remarking that
\begin{align*}
 \langle \boldsymbol{y},  \boldsymbol{\Psi}_{R_{\boldsymbol{u}}(\vartheta,\varphi)} \rangle = T\boldsymbol{y}\left(R_{\boldsymbol{u}}(\vartheta, \varphi)\right).
\end{align*}
concludes the proof.
\item Assume that $\boldsymbol{y} \in \mathbb{R}^{N+1}$.
Since the Kravchuk functions are real-valued,  $\boldsymbol{q}_n \in \mathbb{R}^{N+1}$, and then the coefficients $(\textbf{Q}\boldsymbol{y})[n] = \langle \boldsymbol{y}, \boldsymbol{q}_n\rangle$ are real.
Using that $\forall n \in \lbrace 0,1, \hdots, N\rbrace$, $\mathrm{e}^{2\mathrm{i}\pi n } = 1$, it follows
\begin{align*}
\overline{T\boldsymbol{y}(\vartheta, \varphi)} 
 = &  \sum_{n = 0}^N \sqrt{\binom{N}{n}} \left( \cos \frac{\vartheta}{2} \right)^{n} \left( \sin \frac{\vartheta}{2} \right)^{N - n} \mathrm{e}^{ - \mathrm{i} n \varphi } \overline{(\textbf{Q}\boldsymbol{y})[n]} \\
  = & \sum_{n = 0}^N \sqrt{\binom{N}{n}} \left( \cos \frac{\vartheta}{2} \right)^{n} \left( \sin \frac{\vartheta}{2} \right)^{N - n} \mathrm{e}^{ - \mathrm{i} n \varphi } (\textbf{Q}\boldsymbol{y})[n] \\
= & \sum_{n = 0}^N \sqrt{\binom{N}{n}} \left( \cos \frac{\vartheta}{2} \right)^{n} \left( \sin \frac{\vartheta}{2} \right)^{N - n} \mathrm{e}^{  \mathrm{i} n (2\pi - \varphi) } (\textbf{Q}\boldsymbol{y})[n] \\
= &\,  T\boldsymbol{y}(\vartheta, 2\pi - \varphi).
\end{align*}
Taking the squared modulus on both sides concludes the proof.
\end{enumerate} 
\end{proof}

\section{Multiple frequency components}

\begin{figure*}[t!]
\begin{subfigure}{0.19\linewidth}
\centering
\includegraphics[width = 0.85\linewidth]{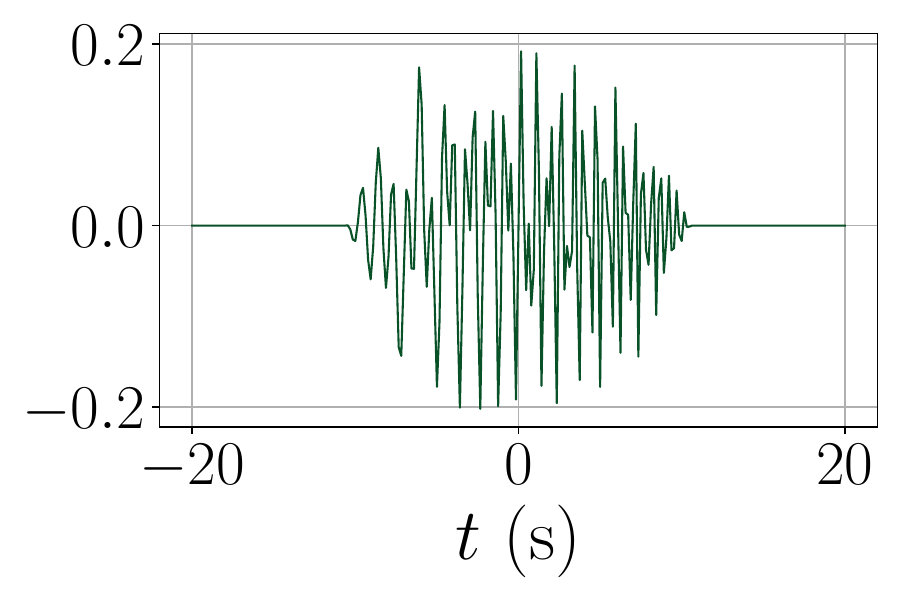}
\end{subfigure}
\begin{subfigure}{0.19\linewidth}
\centering
\includegraphics[width = 0.85\linewidth]{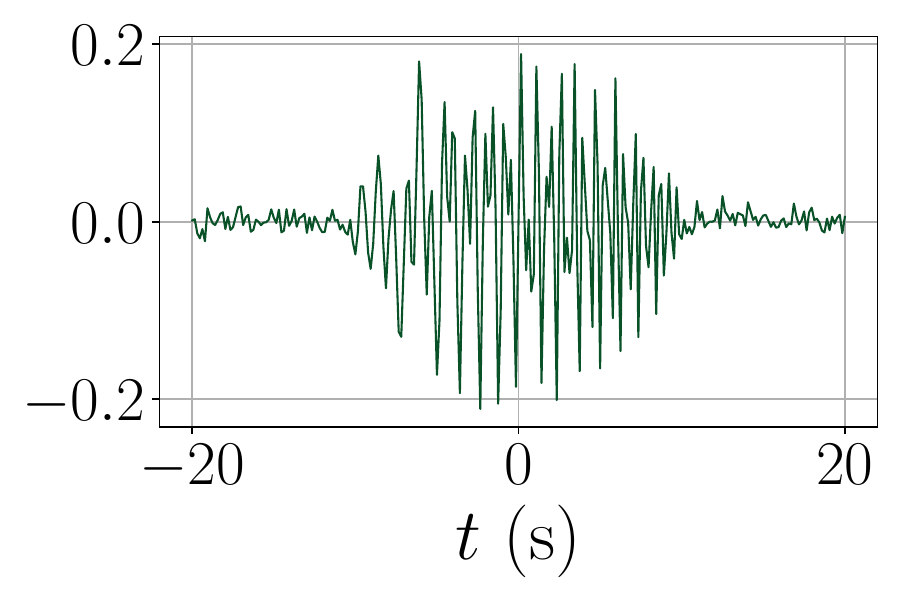}
\end{subfigure}
\begin{subfigure}{0.19\linewidth}
\centering
\includegraphics[width = 0.85\linewidth]{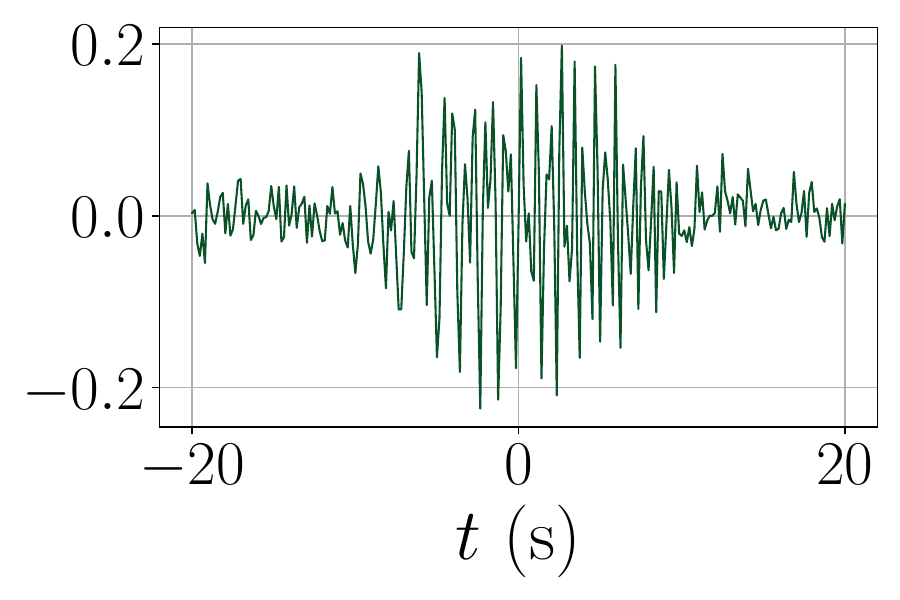}
\end{subfigure}
\begin{subfigure}{0.19\linewidth}
\centering
\includegraphics[width = 0.85\linewidth]{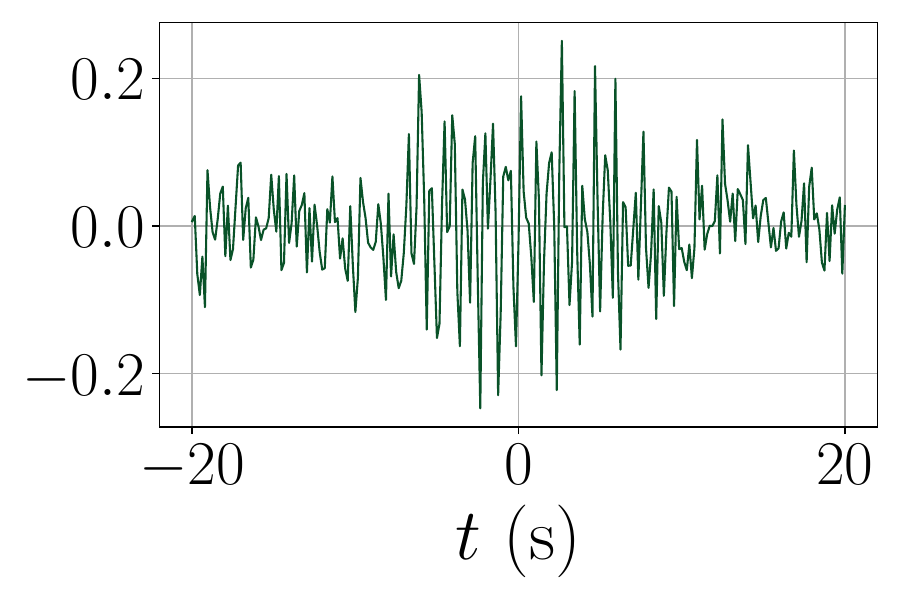}
\end{subfigure}
\begin{subfigure}{0.19\linewidth}
\centering
\includegraphics[width = 0.85\linewidth]{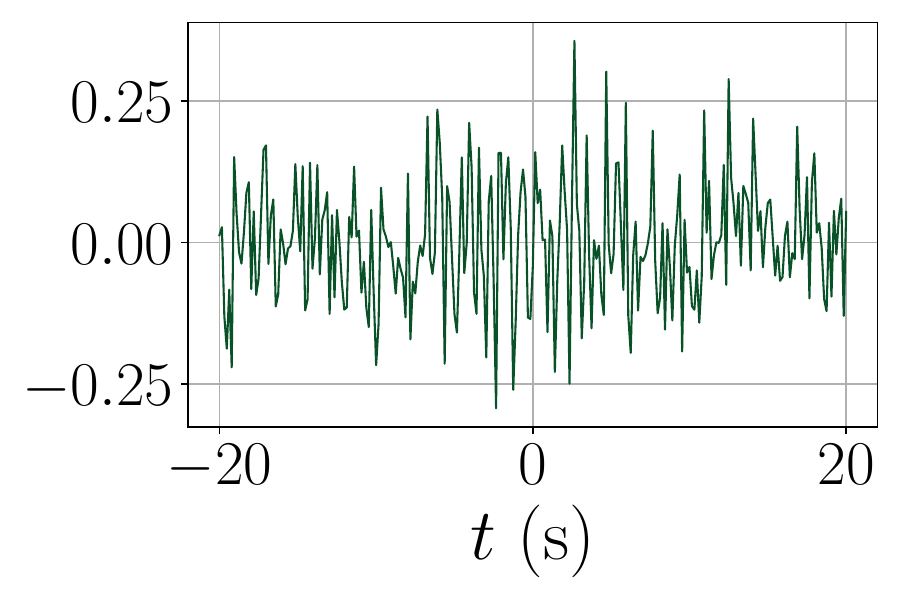}
\end{subfigure}
\begin{subfigure}{0.19\linewidth}
\centering
\includegraphics[width = 0.85\linewidth]{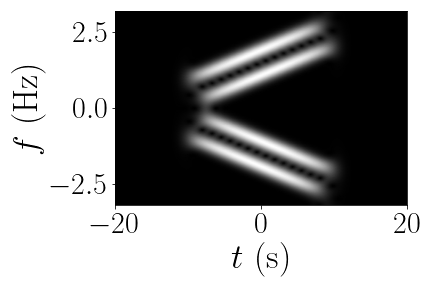}
\end{subfigure}
\begin{subfigure}{0.19\linewidth}
\centering
\includegraphics[width = 0.85\linewidth]{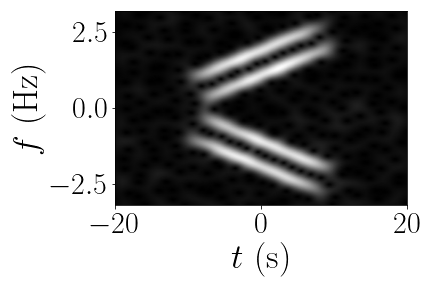}
\end{subfigure}
\begin{subfigure}{0.19\linewidth}
\centering
\includegraphics[width = 0.85\linewidth]{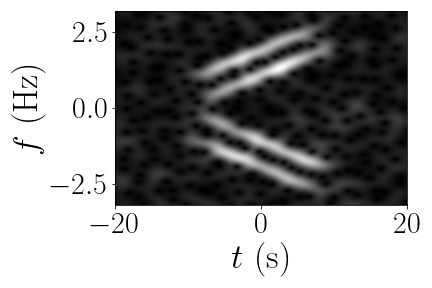}
\end{subfigure}
\begin{subfigure}{0.19\linewidth}
\centering
\includegraphics[width = 0.85\linewidth]{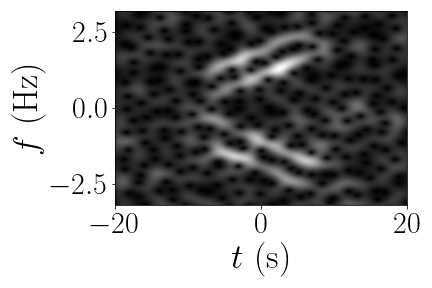}
\end{subfigure}
\begin{subfigure}{0.19\linewidth}
\centering
\includegraphics[width = 0.85\linewidth]{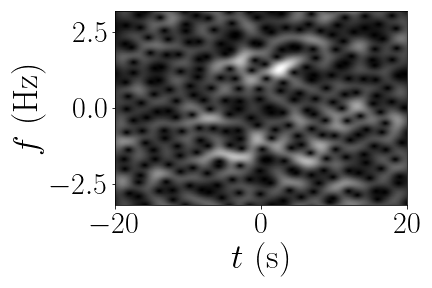}
\end{subfigure}
\begin{subfigure}{0.19\linewidth}
\centering
\includegraphics[width = 0.85\linewidth]{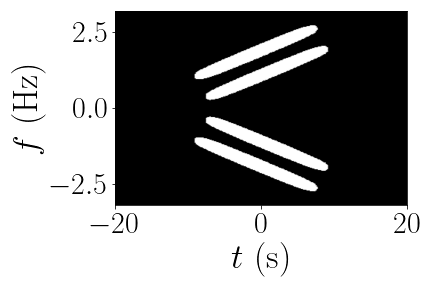}
\end{subfigure}
\begin{subfigure}{0.19\linewidth}
\centering
\includegraphics[width = 0.85\linewidth]{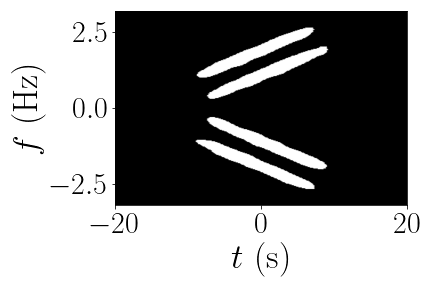}
\end{subfigure}
\begin{subfigure}{0.19\linewidth}
\centering
\includegraphics[width = 0.85\linewidth]{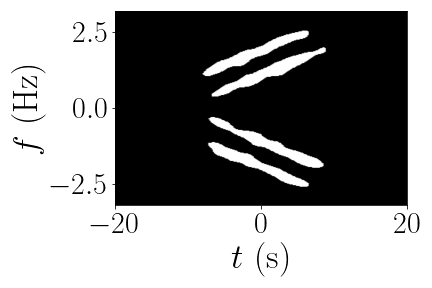}
\end{subfigure}
\begin{subfigure}{0.19\linewidth}
\centering
\includegraphics[width = 0.85\linewidth]{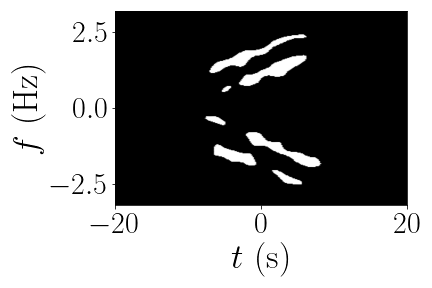}
\end{subfigure}
\begin{subfigure}{0.19\linewidth}
\centering
\includegraphics[width = 0.85\linewidth]{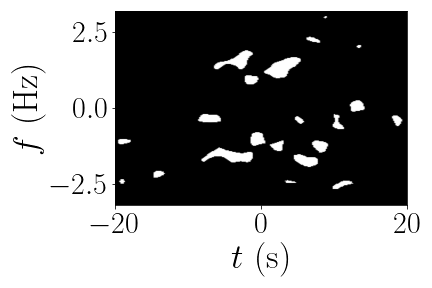}
\end{subfigure}
\begin{subfigure}{0.19\linewidth}
\centering
\includegraphics[width = 0.85\linewidth]{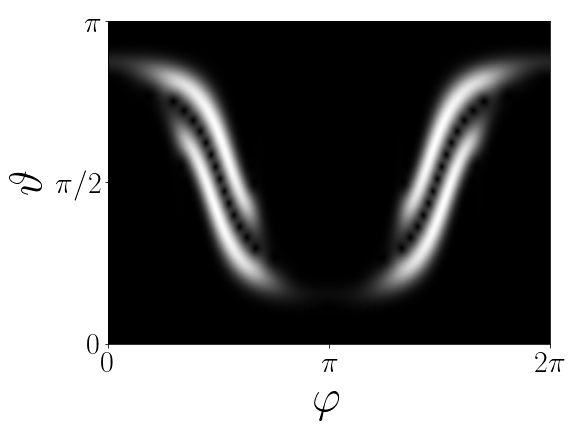}
\end{subfigure}
\begin{subfigure}{0.19\linewidth}
\centering
\includegraphics[width = 0.85\linewidth]{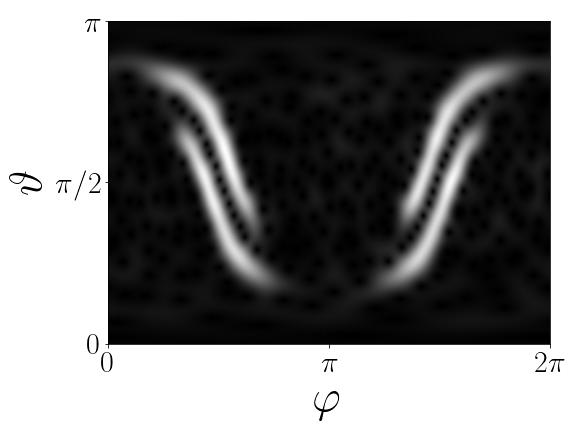}
\end{subfigure}
\begin{subfigure}{0.19\linewidth}
\centering
\includegraphics[width = 0.85\linewidth]{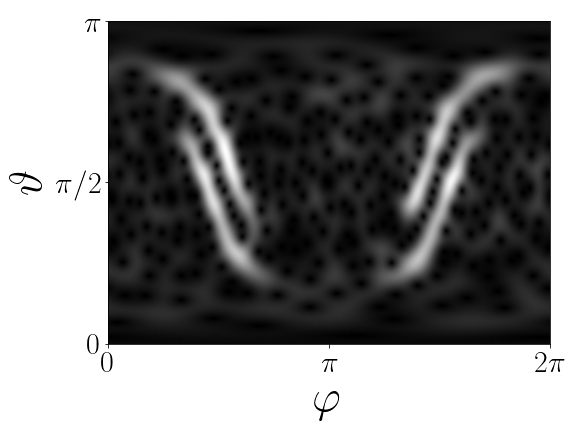}
\end{subfigure}
\begin{subfigure}{0.19\linewidth}
\centering
\includegraphics[width = 0.85\linewidth]{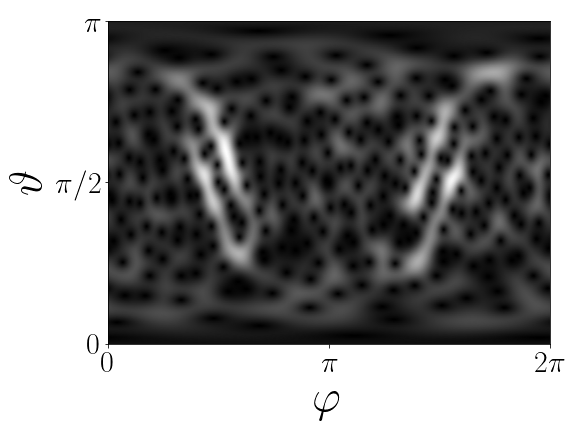}
\end{subfigure}
\begin{subfigure}{0.19\linewidth}
\centering
\includegraphics[width = 0.85\linewidth]{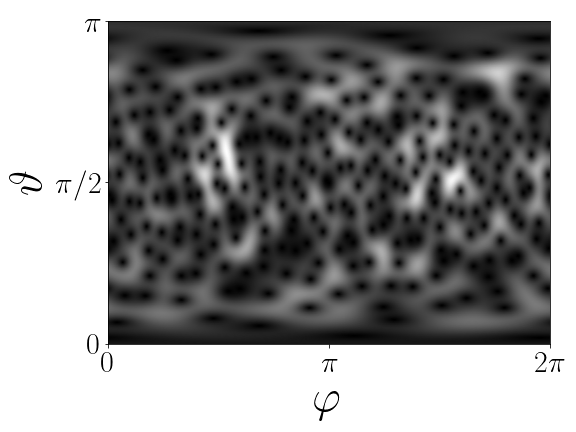}
\end{subfigure}
\begin{subfigure}{0.19\linewidth}
\centering
\includegraphics[width = 0.85\linewidth]{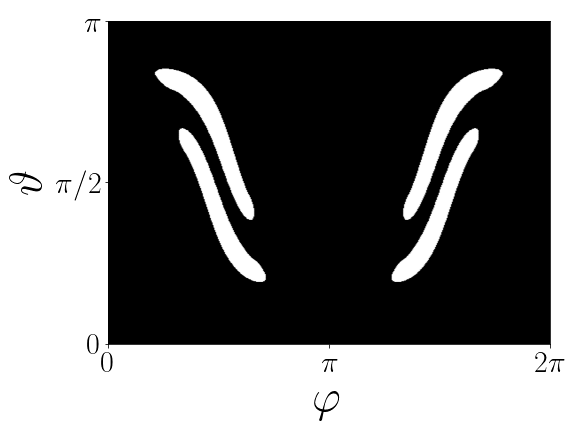}
\subcaption{$\mathsf{snr} = \infty$}
\end{subfigure}\hfill
\begin{subfigure}{0.19\linewidth}
\centering
\includegraphics[width = 0.85\linewidth]{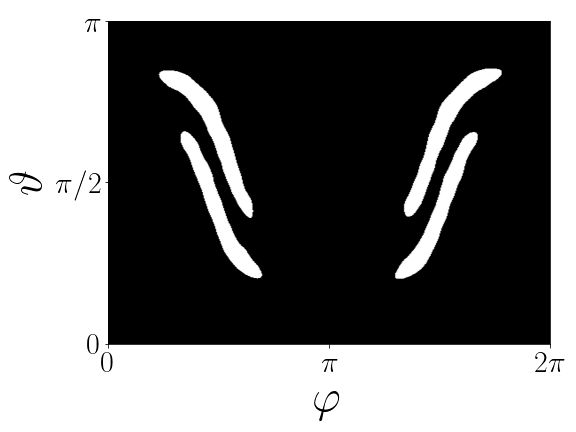}
\subcaption{$\mathsf{snr} = 5$}
\end{subfigure}\hfill
\begin{subfigure}{0.19\linewidth}
\centering
\includegraphics[width = 0.85\linewidth]{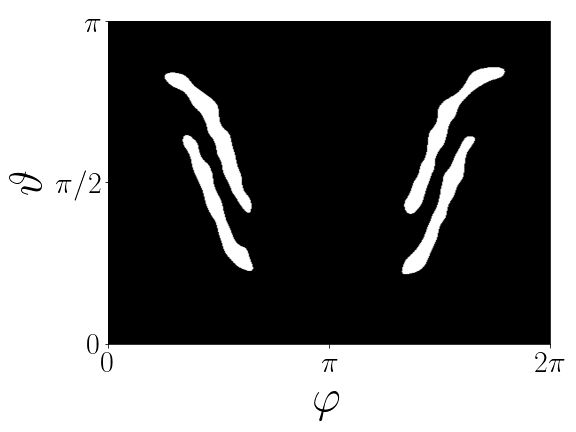}
\subcaption{$\mathsf{snr} = 2$}
\end{subfigure}\hfill
\begin{subfigure}{0.19\linewidth}
\centering
\includegraphics[width = 0.85\linewidth]{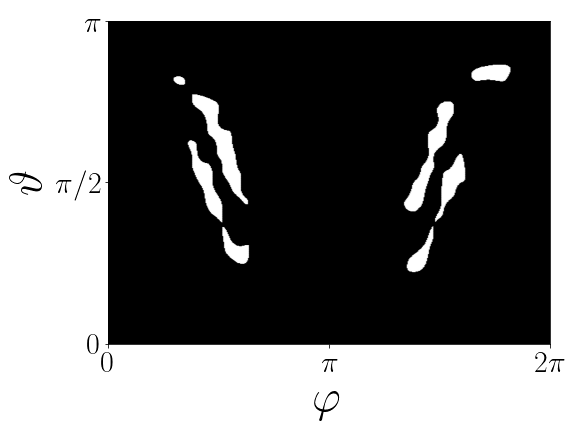}
\subcaption{$\mathsf{snr} = 1$}
\end{subfigure}\hfill
\begin{subfigure}{0.19\linewidth}
\centering
\includegraphics[width = 0.85\linewidth]{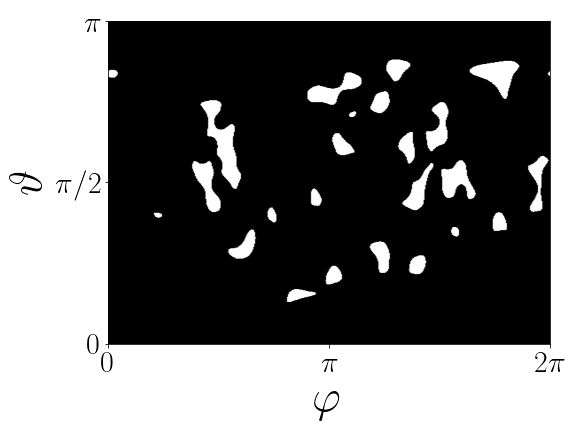}
\subcaption{$\mathsf{snr} = 0.5$}
\end{subfigure}
\caption{\label{fig:component_id}
\textbf{Two-component AM-FM-type signals. } Signals~$\boldsymbol{y}$ follow the AM-FM model~\eqref{eq:AM-FM}, with two components $\boldsymbol{x}_1$ and $\boldsymbol{x}_2$ being two parallel chirps, respectively centered at times $t_1 = -0.75$~s and $t_2 = 0.75$~s and frequencies $f_1 = 1.875$~Hz and $f_2 = 1.125$~Hz. The associated standard Fourier spectrograms and the masks obtained from a $50\%$ amplitude thresholding are respectively displayed in the 2nd and 3rd rows. The Kravchuk spectrograms and the masks obtained from a $50\%$ amplitude thresholding are respectively provided in the 4th and 5th rows. }
\end{figure*}

After tackling the problem of signal detection, the next natural signal processing task to consider would be \textit{component identification}~\cite{flandrin2015time}.
A future research article will be devoted to Kravchuk spectrogram based component identification, disentanglement and denoising, since they raise many interesting and deep theoretical and technical questions about the Kravchuk transform, and notably require a stable implementation of the \textit{inverse} of the Kravchuk transform.
The purpose of this appendix is thus only to provide some illustrations supporting the usability of the Kravchuk spectrogram in various classical signal processing tasks, and in particular for component identification.\\

Given a noisy AM-FM-type signal, with, \textit{e.g.}, two components,
\begin{align}
\label{eq:AM-FM}
\boldsymbol{y} = \mathsf{snr} \times\left( \boldsymbol{x}_1 + \boldsymbol{x}_2\right) + \boldsymbol{\xi}
\end{align}
where $\boldsymbol{x}_1$ and $\boldsymbol{x}_2$ are two chirp signals, $\mathsf{snr}$ is the signal-to-noise ratio and $\boldsymbol{\xi} \sim \mathcal{N}_{\mathbb{C}}(0,\boldsymbol{I})$ is a complex white Gaussian noise, the purpose of \textit{component identification} is to determine the regions $\Omega_i, \, i = 1,2$ of phase space corresponding to each chirp  $\boldsymbol{x}_i$.
Examples of noisy two-component signals, with different noise levels are provided in Figure~\ref{fig:component_id}, first row, where $\boldsymbol{x}_1$ and $\boldsymbol{x}_2$ are two parallel chirps, respectively centered at times $t_1 = -0.75$~s and $t_2 = 0.75$~s and frequencies $f_1 = 1.875$~Hz and $f_2 = 1.125$~Hz.

It appears extremely difficult to devise the two-component character of the underlying signal by direct observation of the temporal signal.
Instead, the classical way to proceed is the turn to a time-frequency representation, \textit{e.g.}, based on the Gaussian spectrogram, as presented in Figure~\ref{fig:component_id}, second row.
A state-of-the-art strategy for component identification then consists in thresholding the spectrogram~\cite{stephane1999wavelet}, as illustrated in Figure~\ref{fig:component_id}, third row.
Similarly, one can devise a component identification strategy by thresholding the Kravchuk spectrograms, provided in Figure~\ref{fig:component_id}, fourth row, which yields the masks presented in Figure~\ref{fig:component_id}, fifth row.\\

Figure~\ref{fig:component_id} shows that the Kravchuk spectrogram identification power is at least as high as that of the Fourier spectrogram, whatever the noise level.
What can be expected for Kravchuk spectrogram is that its identification power is more robust to low sample sizes $N$ compared to that of the Fourier spectrogram.
Further investigations will be the subject of a forthcoming work by the authors, yet Figure~\ref{fig:component_id} is very encouraging for generalizing the use of Kravchuk spectrograms beyond signal detection.

\section{An alterative view on the prefactor in \eqref{e:correspondence_T_L}}
\label{app:moments}

We saw in \eqref{e:correspondence_T_L} that, up to stereographic reparametrization, the Kravchuk transform $T$ is the product of a non-analytic prefactor with the transform $\mathscr{L}$ introduced in \cite{bardenet2021time}. 
While the prefactor naturally appears in the decomposition in spin coherent states that defines $T$, there is another natural way to see why this prefactor naturally turns up if we insist on making $\mathscr{L}$ an isometry. 

The key observation is that binomial coefficients are inverses of moments. 
More precisely, we borrow from \cite[Page 50]{But17} that
\begin{equation}
    (N+1){\binom{N}{k}} = \left( \int_{\mathbb{C}} \vert z\vert^{2k}\e^{-2NV^\nu(z)}\d \nu(z) \right)^{-1},
    \label{e:moment}
\end{equation}
where $\d \nu (z) = \frac{\d z}{\pi (1+\vert z\vert^2)^2}$ is the measure on $\mathbb{C}$ obtained after applying the stereographic projection to the uniform measure on the sphere, and $V^\nu$ is the logarithmic potential of $\nu$, i.e. here $V^\nu(z) = \frac12 \log(1+\vert z\vert^2)$. 
In particular, the density w.r.t. which we take the moments in \eqref{e:moment} is 
$$
g_N(z) \triangleq \e^{-2NV^\nu(z)} = \frac{1}{(1+\vert z\vert^2)^N},
$$
which is the square of the prefactor in \eqref{e:correspondence_T_L}. 
Upon noting this, we can define an isometry from $\mathbb{C}^{N+1}$ to $L^2(\d\nu)$ by defining 
$$
\widetilde{\mathscr{L}} \boldsymbol{y}(z) = \sqrt{(N+1)g_N(z)} \times \mathscr{L} \boldsymbol{y} (z), \quad  \boldsymbol{y}\in\mathbb{C}^{N+1},
$$
so that
\begin{align*}
    \int_{\mathbb{C}} & \vert \widetilde{\mathscr{L}} \boldsymbol{y}(z)\vert^2 \d \nu(z)\\
        &= \int_{\mathbb{C}} \left\vert \sum_{k = 0}^{\infty}\langle \boldsymbol{y}, \boldsymbol{q}_k\rangle \sqrt{\binom{N}{k}} z^k \right\vert^2 (N+1)g_N(z)\d z\\
        &= \sum_{m,n = 0}^{\infty} \langle \boldsymbol{y}, \boldsymbol{q}_m \rangle \overline{\langle \boldsymbol{y}, \boldsymbol{q}_n \rangle}   (N+1) 
        \sqrt{\binom{N}{m} \binom{N}{n}} \int_{\mathbb{C}}  z^{m}\bar{z}^n g_N(z)\d z\\
        &= \sum_{m = 0}^{\infty} \left\vert \langle \boldsymbol{y}, \boldsymbol{q}_m \rangle\right\vert^2\\
        &= \Vert  \boldsymbol{y} \Vert^2.
\end{align*}
Thus $\widetilde{\mathscr{L}}$ preserves norms, and is actually $T$ up to the stereographic mapping. 
It is not clear to us how to justify covariance with a similar argument, though.

\section{Numerical assessment}

For ease of comparison we provide in this appendix the numerical performance reported in Figures~\ref{fig:K_or_F},~\ref{fig:nu_N}~and~\ref{fig:Fourier_or_Kravchuk} in the form of numerical tables.
Each performance measurement (in black) is accompanied with lower and upper bounds of the associated Clopper-Pearson credibility interval at level $0.01$ (in gray).

\begin{table}
\centering
\begin{tabular}{c|cccc}
\toprule
 & \multicolumn{2}{c}{$N+1 = 257$ points}& \multicolumn{2}{c}{$N+1 = 513$ points}\\
 \hline
$\mathsf{snr}$ & Ripley $K$ & Empty space $F$ & Ripley $K$ & Empty space $F$ \\
\hline
$0.5$ 
& $0.05$  {\color{gray} ($0.02$ - $0.1$)} & $0.08$  {\color{gray} ($0.04$ - $0.14$)} 
& $0.07$  {\color{gray} ($0.04$ - $0.10$)}& $0.07${\color{gray} ($0.04$ - $0.10$)} \\ 
$1$ 
& $0.09$ {\color{gray} ($0.05$ - $0.16$)} & $0.15$  {\color{gray} ($0.09$ - $0.22$)} 
& $0.04$ {\color{gray} ($0.02$ - $0.07$)} & $0.26$ {\color{gray} ($0.21$ - $0.31$)}\\
$1.25$ 
& $0.14$ {\color{gray} ($0.08$ - $0.21$)}& $0.28$  {\color{gray} ($0.20$ - $0.37$)} 
& $0.04$ {\color{gray} ($0.02$ - $0.07$)} & $ 0.33$ {\color{gray} ($0.28$ - $0.39$)} \\
$1.5$ 
& $0.18$ {\color{gray} ($0.12$ - $0.26$)}& $0.32$ {\color{gray} ($0.24$ - $0.41$)} 
& $0.11$ {\color{gray} ($0.08$ - $0.15$)} & $0.50$ {\color{gray} ($0.44$ - $0.56$)}\\
$2$ 
& $0.28$ {\color{gray} ($0.20$ - $0.36$)}& $0.66$ {\color{gray} ($0.56$ - $0.74$)} 
& $0.09$ {\color{gray} ($0.06$ - $0.30$)} & $ 0.68$ {\color{gray} ($0.62$ - $0.73$)} \\
$5$ 
& $0.93$ {\color{gray} ($0.87$ - $0.97$)}& $1.00$ {\color{gray} ($0.97$ - $1.00$)} 
& $0.25$ {\color{gray} ($0.20$ - $0.30$)}  & $1.00$ {\color{gray} ($0.99$ - $1.00$)} \\
$10$ 
& $1.00$ {\color{gray} ($0.97$ - $1.00$)} & $1.00$ {\color{gray} ($0.97$ - $1.00$)} 
& $0.64$ {\color{gray} ($0.58$ - $0.69$)} & $1.00$ {\color{gray} ($0.99$ - $1.00$)} \\
$50$ 
& $1.00$ {\color{gray} ($0.97$ - $1.00$)}& $1.00$ {\color{gray} ($0.97$ - $1.00$)}
& $1.00$ {\color{gray} ($0.99$ - $1.00$)}& $1.00$ {\color{gray} ($0.99$ - $1.00$)}\\
\bottomrule
\end{tabular}
\caption{\label{tab:K_or_F} \textbf{Comparison between $K$ and $F$ functional statistics.}
Evolution of the power of the test with the signal-to-noise ratio (in black) with Clopper-Pearson confidence intervals at level $0.01$ (lower and upper bounds in gray).  \textit{(Numerical results presented in Figure~\ref{fig:K_or_F})}.}
\end{table}

\begin{table}
\centering
\begin{tabular}{c|cccc}
\toprule
 & \multicolumn{2}{c}{$\mathsf{snr} = 2$}& \multicolumn{2}{c}{$\mathsf{snr} = 1.5$}\\
 \hline
$N$ & $2 \nu = 30$~s & $2\nu = 20$~s & $2\nu = 30$~s & $2\nu = 20$~s \\
\hline
$128$
&  $0.56$ {\color{gray} ($0.47$ - $0.65$)} & $0.47$ {\color{gray} ($0.39$ - $0.54$)}
&  $0.33$ {\color{gray} ($0.25$ - $0.42$)} & $0.24$ {\color{gray} ($0.18$ - $0.30$)}\\
$256$ 
&  $0.66$ {\color{gray} ($0.56$ - $0.74$)} & $0.36$ {\color{gray} ($0.29$ - $0.43$)} 
&  $0.32$ {\color{gray} ($0.24$ - $0.41$)} & $0.24$ {\color{gray} ($0.18$ - $0.31$)}\\
$512$ 
&  $0.68$ {\color{gray} ($0.62$ - $0.73$)} & $0.25$ {\color{gray} ($0.19$ - $0.32$)}
&  $0.50$ {\color{gray} ($0.44$ - $0.56$)} & $0.29$ {\color{gray} ($0.22$ - $0.35$)}\\
$1024$
&  $0.89$ {\color{gray} ($0.82$ - $0.94$)} & $0.51$ {\color{gray} ($0.44$ - $0.58$)}
&  $0.68$ {\color{gray} ($0.58$ - $0.76$)} & $0.34$ {\color{gray} ($0.28$ - $0.41$)}\\
\bottomrule
\end{tabular}
\caption{\label{tab:nu_N} \textbf{Robustness to small number of samples and short duration.}
Evolution of the power of the test with the length of the observation (in black) with Clopper-Pearson confidence intervals at level $0.01$ (lower and upper bounds in gray).
\textit{(Numerical results presented in Figure~\ref{fig:nu_N})}.}
\end{table}

\begin{table}
\centering
\begin{tabular}{c|cccc}
\toprule
 & \multicolumn{2}{c}{$2\nu = 30$~s}& \multicolumn{2}{c}{$2\nu = 20$~s}\\
 \hline
$N$ & Fourier & Kravchuk &   Fourier & Kravchuk \\
\hline
$128$
& $0.02$ {\color{gray} ($0.00$ - $0.06$)} & $0.33$ {\color{gray} ($0.25$ - $0.42$)}
&  $0.04$ {\color{gray} ($0.01$ - $0.08$)} & $0.24$ {\color{gray} ($0.18$ - $0.30$)}\\
$256$ 
& $0.03$ {\color{gray} ($0.01$ - $0.07$)} & $0.32$ {\color{gray} ($0.24$ - $0.41$)}
& $0.04$ {\color{gray} ($0.01$ - $0.09$)}& $0.24$ {\color{gray} ($0.18$ - $0.31$)}\\
$512$
& $0.06$ {\color{gray} ($0.03$ - $0.12$)} & $0.50$ {\color{gray} ($0.44$ - $0.56$)}
& $0.06$ {\color{gray} ($0.02$ - $0.11$)}& $0.29$ {\color{gray} ($0.22$ - $0.35$)}\\
$1024$
& $0.32$ {\color{gray} ($0.24$ - $0.41$)} & $0.68$ {\color{gray} ($0.58$ - $0.76$)}
& $0.22$ {\color{gray} ($0.15$ - $0.30$)}& $0.34$ {\color{gray} ($0.28$ - $0.41$)}\\
\bottomrule
\end{tabular}
\caption{\label{tab:Fourier_or_Kravchuk} \textbf{Detection tests based on the zeros of either Fourier or Kravchuk spectrogram.}
Evolution of the power of the test with the signal length $N$ for noisy signals with fixed $\mathsf{snr} = 1.5$ (in black) with Clopper-Pearson confidence intervals at level $0.01$ (lower and upper bounds in gray).
\textit{(Numerical results presented in Figure~\ref{fig:Fourier_or_Kravchuk})}.}
\end{table}

\bibliographystyle{plain}
\bibliography{my_biblio,stats}

\end{document}